\documentclass[showpacs,aps,singlecolumn]{revtex4-2}
\usepackage{graphicx}
\usepackage{amsmath,amssymb,amsfonts}
\usepackage{array}
\usepackage{url}
\usepackage{hyperref}
\usepackage{multirow}
\usepackage{float}
\usepackage{lineno}
\usepackage{xspace}
\usepackage{ulem}
\usepackage{orcidlink}
\begin{document}

\title{QCD Axion Conversion in Magnetospheres of Neutron Stars}
\author{Shubham Yadav\,\orcidlink{0000-0002-6716-7804}}
\email{shubhamyadav@oriental.ac.in}
\affiliation{Department of Basic Sciences, Oriental Institute of Science and Technology, Bhopal, Madhya Pradesh, India}

\author{Sachin Dev Kushwaha\,\orcidlink{0009-0008-1936-6551}}
\email{sachinkh15@gmail.com}
\affiliation{Department of Basic Sciences, Oriental Institute of Science and Technology, Bhopal, Madhya Pradesh, India}

\author{Kamni Pawar\,\orcidlink{0009-0003-2064-353X}}
\email{kamnipawar@oriental.ac.in}
\affiliation{Department of Basic Sciences, Oriental Institute of Science and Technology, Bhopal, Madhya Pradesh, India}

%\author{Tapomoy Guha Sarkar}
%\email{tapomoy@pilani.bits-pilani.ac.in}

%\affiliation{Department of Physics, Birla Institute of Technology and Science, Pilani, Rajasthan, India}

%\author{Captain R. Singh}
%\email{captainriturajsingh@gmail.com}
%\affiliation{Department of Physics, Indian Institute of Technology Indore, Simrol, Indore 453552, India}

\begin{abstract}
The axion-converted-photons flux plays a key role in searches of QCD axions as a dark matter (DM) candidate. In addition to solving the strong CP problem, these may explain the properties
of the mysterious DM. Neutron star (NS) cooling by neutrino/axion emissions rate constrains the astrophysical properties of superdense matter. We attempt to analyse the impact of strong magnetic fields on the emission properties of axions by employing the APR, FPS, and SLY equation of State (EoS). We modify the Tolman Oppenheimer Volkoff (TOV) equations by considering effects of strong fields' and generate profiles. We assume Cooper-pair-breaking formation (PBF) and the Bremsstrahlung process occurring in the core of NS. We use the radial distance-dependent magnetic field configuration within the permissible limit of the central magnetic field. Our work explores the variations in the axion masses of 10 meV, 15 meV, and 20 meV as a cooling source for spherically symmetric magnetised NSs. We have also shown the energy spectrum of axions and their conversion to photons. Our analysis shows that the cooling rate and the luminosity of observables (axions) for NSs change significantly due to the magnetic field. The results for the luminosity of axions is quite sensitive to its mass and EoS, as the less separation is observed for SLY EoS in comparison to the FPS and APR EoS. We report that the spectrum of axions from the Bremsstrahlung process dominates over the PBF process at lesser axion energies, within the possible axion mass range for PSR J1357-6429 NS. Our results reveal that the impact of the magnetic field is less at lower axion energies, indicating the necessity for including a magnetic field in axion-to-photon conversion mechanisms. The results for the axions mass 10 meV and 15 meV as compared to 20 meV confirms the existing report of the 16 meV.

\end{abstract}
\date{\today}
\maketitle

\section{Introduction}
\label{intro}
Dark matter (DM) is one of the Universe's most challenging mysteries, which is yet to be solved. A
considerable portion of the Universe is made up of an unknown non-luminous matter that
varies fundamentally from conventional forms of matter, accounting for $5\%$ of the
Universe's visible matter, such as planets, stars, humans, and galaxies. The $25\%$ Universe
is made up of an unexplained form of matter called DM. At the moment of decoupling,
the DM particles' density and the Universe's temperature determine whether these
particles are relativistic or non-relativistic in nature. These will be relativistic when their
mass is greater than the temperature of the Universe and are referred to as hot DM
candidates. In contrast, those with non-relativistic velocities are called cold DM candidates. These
are warm DM particles when the velocities lie between hot and cold DM. Annihilations in
the galaxy lead to the relative production of elementary particles like photons, neutrinos,
and cosmic rays. Some crucial classes related to detection, including the annihilation rate,
DM mass, and scattering rate of matter, significantly impact the signals received on
Earth~\cite{Pshirkov_2009}. These signals and the detectors that detect neutrinos and photons could probe the
identity of the mysterious substance giving some possibility of indirect detection of DM, especially when it is assumed to be a weakly interacting particle. Various DM candidates
have been proposed in the literature, among which QCD axions or Axion-like particles
(ALPs) are believed to be the most promising candidates for the searches.
QCD Axions emerged as a solution to the strong CP problem in particle physics. Being less massive and lesser energy, these particles are postulated to exist due to some properties related to strong force~\cite{Buschmann:2019icd,chadha2022axion,PhysRevLett.40.279,ABBOTT1983133,PRESKILL1983127,DINE1983137,PhysRevD.32.3172,PhysRevLett.53.1198,PhysRevLett.120.151301,yadav2022emission}. The recent experimental observations/findings have completely surpassed the initial assumption of strong interaction of axupper bound ofions' with the matter. Various limits on axion mass have also been presented in the literature~\cite{PhysRevLett.117.141801,PhysRevD.97.123006,PhysRevD.92.075012,PhysRevLett.123.021801,Marsh_2017}. The axion properties are strongly dependent on its mass, which is inversely proportional to the decay constant $f_{a}$ given as~\cite{PhysRevD.34.843,PhysRevD.37.1237,PhysRevLett.123.061104}:
\begin{equation}
m_{a}=0.6 eV \frac{10^{7}GeV}{f_{a}}
\end{equation}
Currently,the study for axion physics is based on two models: the
Dean-Fischler-Srednitsky-Zhitnitsky (DFSZ) and Kim-Shifman-Weinstein-Zakharov
(KSVZ) model~\cite{Leinson_2019,Duffy_2009,Peccei:1977hh,Peccei:1977ur}. 
Various neutrino/axion production mechanisms have been proposed in the literature~\cite{PhysRevLett.66.2701,2001PhR...354....1Y,PhysRevLett.66.2701,Leinson_2000}.
To understand the physics behind these weakly interacting particles one needs to understand the NS cooling~\cite{Beznogov_2023,Page_2004,Yakovlev_2005,YAKOVLEV2004523,PAGE2006497,PhysRevD.37.2042,refId0,Buschmann:2021juv,Kaminker_2006,Valyavin2014SuppressionOC,Page:2005fq,wijnands2017cooling,PRAKASH1994297,PhysRevLett.106.081101}. NSs serves as an excellent astrophysical lab,
allowing us to explore the observable properties, structure and composition of matter at
low temperature and high-density~\cite{PRAKASH19971,Lattimer_2001,2006NuPhA777497P,2006ARNPS56327P,2004Natur.431..819B,rathod2024cooling}. As the stars get out of the fuel (in the last stage of life span), a core-collapse supernova explosion occurs, and the leftover core
leads to the formation of compact objects like NSs, and white dwarfs. Observations from
the Soft Gamma-Ray Repeaters (SGRs) and X-ray dim isolated NSs (XDINS) have
revealed information regarding the existence of strong magnetic fields in their interiors and
magnetospheres~\cite{DEXHEIMER2017487,Dexheimer:2017fhy,Lopes_2015,2013arXiv1307.7450G,10.1111/j.1365-2966.2008.14034.x,1995A&A...301..757B,Geppert_2006,dexheimer2012hybrid,Gomes_2017}. The stars with such strong magnetic fields are Magnetars. Various theories have been proposed to study the macroscopic structure and thermodynamic behavior of the strongly magnetized NSs~\cite{2021EPJC...81..698P,Haensel_2002,SINHA201343,Sedrakian:2006mq,2006A&A...450.1077B,yadav2024observed,yadav2023axion}. The superdense matter's interior composition and properties depend on the employed equation of state (EoS) (relation
between energy density and pressure)~\cite{Gusakov_2005,Schneider_2019,unknown,PhysRevC.58.1804,Broderick_2000,yadav2024thermal}. The first research started with the
concept of relativistic Lagrangian and the Hartree approximation, used for
calculating the EoS with all possible occupied baryon states of the superdense medium.
The $\pi$ contribution and the $\rho$ meson's tensor coupling, along with the incompressibility of
nuclear matter that does not reconcile with $\sigma$ meson self-interactions, was not included in
the research. The EoS also plays a vital role in our knowledge of core-collapse
supernovae and many other astronomical processes, including mergers of compact stars
or cooling proto-neutron stars~\cite{Kolomeitsev_2008,Potekhin_2020,Mignani_2013,Dessert_2022,10.1093/mnras/sty776,10.1093/mnras/stx366}. 
Paul et al.~\cite{Paul2018NeutronSC} investigate the nature of NSs cooling by calculating the axion energy loss
rate from Bremsstrahlung process under the degenerate and non-degenerate limits and
compare the findings with the assumption that cooling is solely caused by emission of
gamma rays and neutrinos. Pshirkov et al.~\cite{Pshirkov_2009} provide a novel technique to identify DM
axions that uses radio measurements by observations from magnetospheres of NSs. This
provides a strong foundation for the conversion of axions to photons in an intense
magnetic field of NSs. Millar et al.~\cite{millar2021axion} first time provides the numerical simulation
results of axion to photon conversion for a highly magnetized anisotropic material in full 3-Dimensions.
In this work, we present the effect of magnetic fields on the energy spectrum of
axions and the conversion of axions to photons in magnetospheres of the NSs. It has been
assumed that the NSs magnetospheres and interiors hold a strong magnetic field. We start from solving the TOV equations in the presence of magnetic field and generate profiles. By employing APR, FPS and SLY EoS~\cite{PhysRevC.58.1804} and profiles, we derive the results for cooling and luminosity as a function of the characteristic age at 10 meV, 15 meV and 20 meV axion mass. We focus on PSR J1357-6429 with the characteristic age of $7 \times 10^3$ years, and show the maximum departure for the magnetic field as compared to the no-magnetic field case. The paper is structured as follows: Section 1 gives a short introduction. In section 2, we present details on modified TOV equations, NS cooling, neutrino and axion emission rate, energy
spectrum of axions, and conversion probability. Section 3 presents the Results and
Discussion of the existing research. Finally, section 4 summarizes the study and discusses
the significant findings with the future scope.

\section{Formalism}
\label{form}

The current calculations are based on the solutions for set of TOV equations in the presence of the
magnetic field. The TOV equations are first-order differential equations based on the solution of Einstein's general relativity with the metric in the interior of a star~\cite{PhysRev.55.364,PhysRev.55.374}. Under the influence of strong magnetic fields a magnetic energy density term is added to the matter-energy density, thereby making TOV equations in the presence of a magnetic field. These strong fields will influence the internal
solutions by acting as a source of the energy-momentum tensor. Furthermore, the effect
of the magnetic field can be studied by including a Lorentz force term $\mathcal L$ in the pressure
gradient equation under the broad relativistic frameworks for conserving the energy
momentum tensor. For stability, a symmetric NS should have both toroidal and poloidal components in the same ratio.
We have considered the radial distance-dependent magnetic field $B(r)$, which depends on the central magnetic field $B_{c}$ given by~\cite{10.1093/mnras/stu2706,2019PhRvC..99e5811C}:
\begin{equation}
B_0(r) = B_c\left [ 1 - 1.6 \left(y \right )^2  -  \left(y \right )^4 + 4.2  \left(y \right )^6 -2.4  \left(y \right )^8 \right],
\end{equation}
Here $y=\frac{r}{\bar r}$ and $\bar r$ represents distance slightly greater than the actual star`s radius. The validity of the profile for PSR J1357-6429 relies on its thermal evolution, cooling behaviour, and observations from XMM-Newton and Chandra data sets. In the presence of a magnetic field, the TOV equations for mass and hydrostatic equilibrium are given by~\cite{2019PhRvC..99e5811C}:
\begin{equation}
\frac{dm}{dr}=4\pi r^{2}\left ( \epsilon +\frac{B^{2}}{2\mu _{0}c^2} \right )
\end{equation}

\begin{equation}
\frac{dP}{dr}=-c^{2} \left ( \epsilon +\frac{B^2}{2\mu _{0}c^2}+\frac{P}{c^{2}} \right )\left ( \frac{d\phi }{dr}-{\mathcal L}\left ( r \right ) \right ),
\end{equation}
Here $\frac {d\phi }{dr}$ is the first order differential equation of gravitational potential. Under the magnetic field, an effective Lorentz force term is incorporated in the modified TOV equations given as~\cite{2019PhRvC..99e5811C}:
\begin{equation}
\mathcal L\left ( r \right )  = B_{c}^2\left[-3.8 \left(y \right ) + 8.1 \left(y \right )^3  -1.6   \left(y \right )^5 -2.3   \left(y \right )^7 \right ]\times 10^{-41}
\end{equation}
Solving the TOV equations numerically with an EoS we get the variation of global quantities
i.e. mass, gravitational potential, and hydrostatic equilibrium as a function of the radial distance (profiles).  
To study the effects of strong magnetic fields on the cooling and the state of highly dense matter of NSs, we use the NSCool code~\cite{2016ascl.soft09009P}. This Fortran-77 code features robust, fast, and modular software that solves the equations under a full general-relativistic framework. The solutions to the equations for transport of heat and conservation of energy, the profiles for the global quantities and the EoS is an integral input to code. The traditional 1-dimensional version handles several standard configuration files. Some of them exist as data files and parameters are linked with the subroutine files. This results in a series of temperature profiles (as a function of time and distance) and luminosity of observables (photons and neutrinos) as a function of the characteristic age~\cite{Page_2009,refId02,2019A&A...629A..88P,2015MNRAS.447.1598B}. For current work the 1.3 $M\odot$ NS stellar mass profiles and three EoS is an input to several built-in structural and physical parameters of the NSCool code. The robustness and accuracy of code, is done using the specific tests related to self-convergence and a comparative study with the several analytical constraints.
Under the Newtonian framework the equation for the conservation of energy becomes

\begin{equation}
C_{v}\frac{\mathrm{dT_b^{\infty}} }{\mathrm{d} t}=-L_{\nu }^\infty-L_{\gamma }^\infty(T_{s})+H,
\end{equation}

Here $L_{\nu}$ and $L_{\gamma}$ is neutrino and photon luminosity respectively and H is heating mechanism. Several heating mechanisms are present in the code but we have not included in the current calculations.
To study the physics behind the axions we have added the expression for its emissivity in the NSCool code~\cite{Buschmann:2019pfp}. We have assumed that the axions along with neutrinos are emitted from the core of the NSs by the Cooper-pair-breaking formation (PBF) and Bremsstrahlung process. The states of superfluidity and relative gap models are not included in current mechanisms for axion production. Magnetic fields have strong imprints on the thermal conduction in the outer layers of the NS, thereby affecting heat flows close to the stellar surface and the actual surface temperature. The heat blanketing layer is present in the outermost part of the core within the distance of $\sim 100 m$ of the NS. Additionally, the thermal conductivity of electrons can be significantly impacted by a magnetic field within the heat blanketing layer~\cite{2012A&A...538A.115P,refId04,refId01,Potekhin_2003,2016MNRAS.459.1569B}.
The expression for the neutrino and axion emission rate are adapted from the Refs~\cite{Keller_2013,Sedrakian_2019,Sedrakian_2016}:
The first order derivative of emissivity with respect to local axion energy is the energy spectrum of axions given as~\cite{Buschmann:2019pfp}:

\begin{equation}
{E_{ax}}= \frac{d\epsilon_{a}}{d\omega_{a}}, 
\end{equation} 

The energy spectrum of axions from the S-wave (spin $0$) pairing:

\begin{equation}
{E_{ax}^{s}}= \frac{Norm^{s}}{2\Delta T}\frac{\left ( \frac{\omega_{a} }{2\Delta T} \right )^3}{\sqrt{\left (\frac{\omega_{a} }{2\Delta T} \right )^2-1}}\left [ f_{F}(\frac{\omega_{a}}{2T})\right ]^2,
\end{equation}
where,
$$
{Norm^{s}} = \epsilon_{a}^{s}~z_{n}^{5}/I_{a}^{s}
$$
is normalization constant calculated from the equation $\int_{2\Delta T}^{\infty }E_{ax}^{s}d\omega_{a}$ and  $2y\Delta T$ is the energy of axion.

The energy spectrum of axions from the P-wave (spin $1$) pairing:

\begin{equation}
{E_{ax}^{p}}= \frac{d\epsilon_{a}^{p}}{d\omega_{a}},
\end{equation}

\begin{multline*}
{E_{ax}^{p}} =  \int_{-1}^{1}dcos\theta\frac{1}{4}\Delta^{P}(T,\theta)^4 {Norm^{p}}\\ \times \frac{\left ( \frac{\omega_{a} }{2\Delta^{P} (T,\theta)} \right )^3}{\sqrt{\left (\frac{\omega_{a} }{2\Delta^{P} (T,\theta)} \right )^2-1}}\left [ f_{F}(\frac{\omega_{a}}{2T})\right ]^2,
\end{multline*}

$$
{Norm^{p}} = \epsilon_{a}^{p}/T^{5}~I_{a}^{p}
$$
is the normalisation constant which is defined as $\int_{2\Delta^{P}(T,\theta)}^{\infty }E_{ax}^{p}d\omega_{a}$ and 2y$\Delta^{P}(T,\theta)$ is the axion energy $\omega_{a}$. 

The equation for the energy spectrum of axions from N-N bremsstrahlung process is given as:
\begin{equation}
{E_{ax}^{br}}=\frac{d \epsilon_{a}}{d\omega_{a}} = \frac {Norm^{br}\omega_{a}^{5}} {e^{x} T^6},
\end{equation}
where, 
$Norm^{br}= \frac{\epsilon_{a}}{4 \pi^{2} I_{a}}$ is normalization constant. This is derived by $\int_{0}^{\infty }E_{ax}d\omega_{a} =\epsilon_{a}$.
Here, $e^{x}=e^{\frac{\omega_{a}}{T}}$ is the dimensionless number and $\omega_{a}$ is energy of axion.

Axions produced in the cores get converted to photons due to the presence of intense magnetic fields in the magnetospheres of NSs. 
The expression of the conversion probability (axions to photons) is given by~\cite{Buschmann:2019pfp}:
\begin{equation}
\begin{split}
P_{a\to\gamma }\approx 1.5\times 10^{-4}\left(\frac{g_{a\gamma\gamma}}{10^{-11}\,GeV^{-1}}\right)^2\left(\frac{1\,keV}{\omega}\right)^{0.8} \\
\times\left(\frac{B_{0}}{10^{13}\,G}\right)^{0.4}\left(\frac{R_{NS}}{10\, km}\right )^{1.2}\sin^{0.4}\theta,
\end{split}
\end{equation}

$g_{a\gamma \gamma}$ is axion-photon mixing term, $R_{NS}$ is the radius of NS, B is magnetic
field and $\theta$ is the angle from the magnetic axis. Finally, multiplication of the energy
spectrum of axions with the conversion probability gives the axion-converted-photon
flux. The formula does not include resonant conversion effect (Landau-Zener formula), thereby weakens its validilty for axion masses in the range probed for NS. It presents an approximate analysis in comparison to Goldreich-Julian plasma model.

\section{Results and Discussions}
\label{rds}
The current analysis is primarily carried out to study the influence of strong magnetic fields on the non-magnificent seven PSR J1357-6429 NS (different from the well-known group of M7 stars), using the NSCool code. This section addresses the effect of different axion masses on key components within the framework of NS cooling, obtained by assuming that axions production mechanisms (PBF and Bremsstrahlung process) occurs within the core of NS. 

%%%%%%%%%%%%%%%%%%%%%%%%%%%%%%%%%%%%% TvsR %%%%%%%%%%%%%%%%%%%%%%%%%%%%%%%
%%%%%%%%%%%%%%%%%%%%%%%%%%%%%%%%%%%%%%%%
\begin{figure*}[htp!]
\begin{tabular}{ccc}
\includegraphics[width=5.55cm]{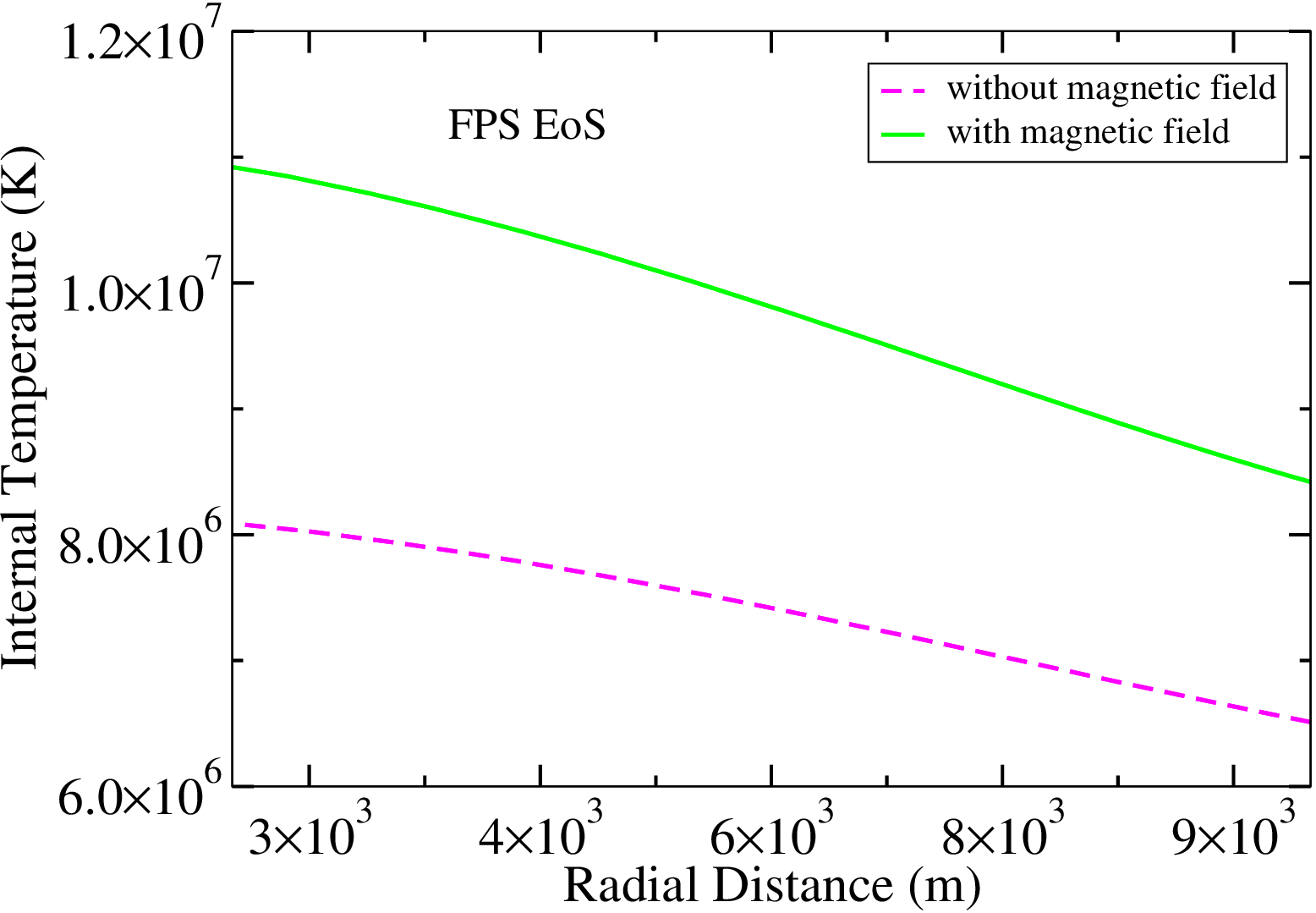}\ & \includegraphics[width=5.55cm]{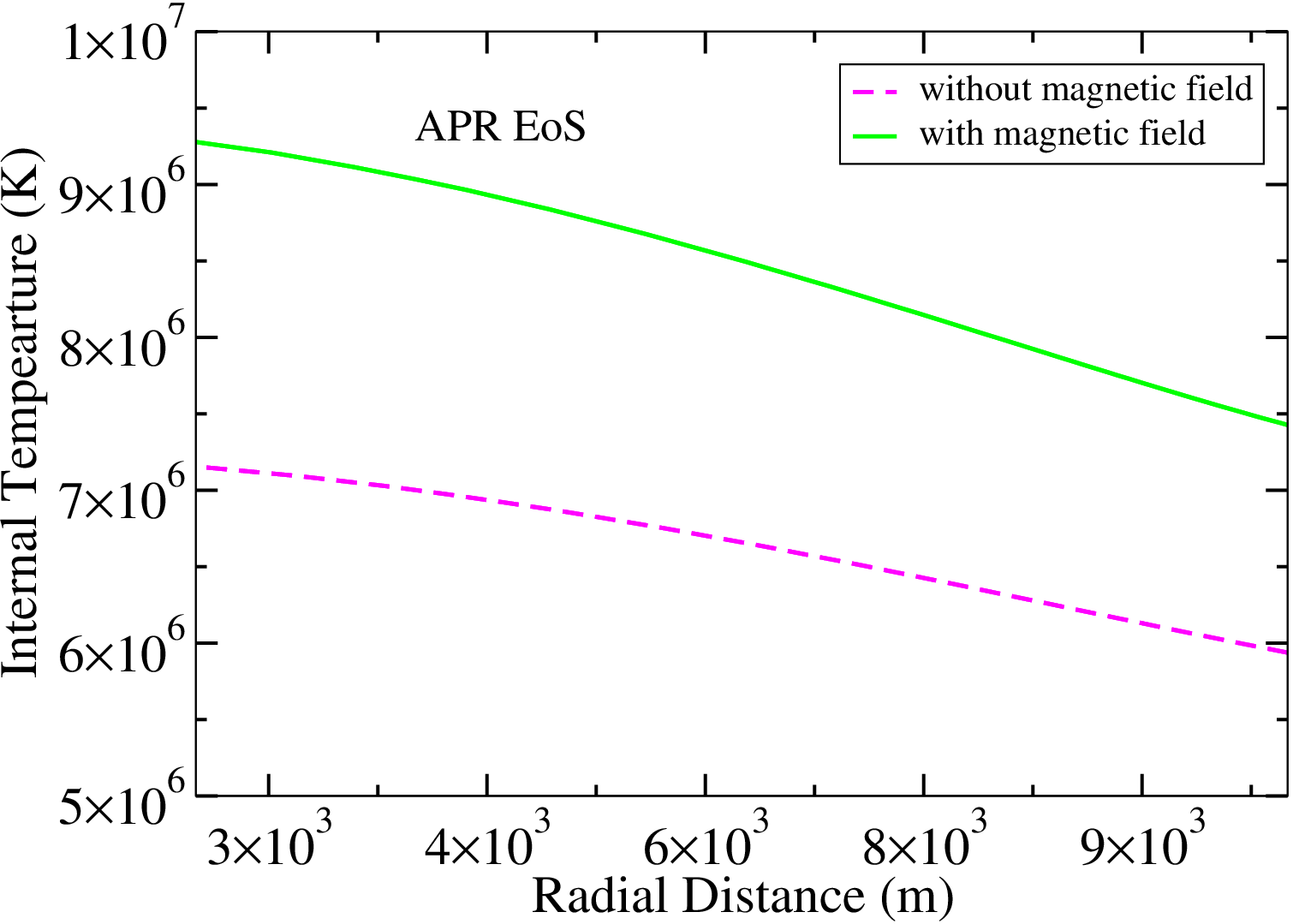}\ & \includegraphics[width=5.55cm]{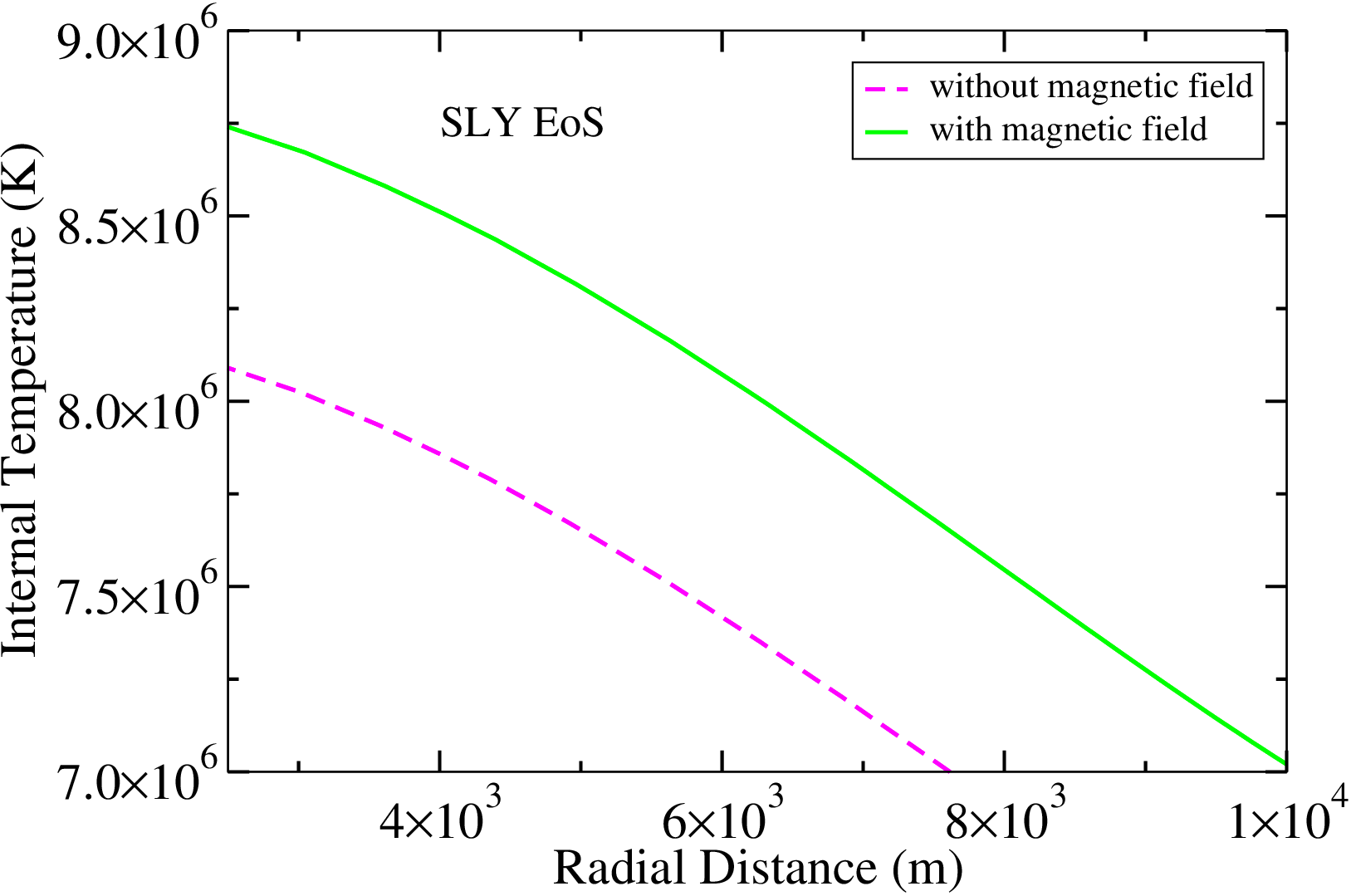} 
\end{tabular}
\caption{The internal temperature distribution versus radial distance (R) for without and with magnetic field at axion mass of 10 meV.} 
\label{fig:nstr1} 
\end{figure*}
%%%%%%%%%%%%%%%%%%%%%%%%%%%%%%%%%%%%%%%
Figure (\ref{fig:nstr1}) analyses the impact of EoS (FPS, APR, and SLY EoS) at an axion mass of 10 meV. It represents the changes in internal temperature as a function of radial distance, with and without a magnetic field. It indicates that the cooling behaviour changes significantly under the effect of the strong magnetic field.
\begin{figure*}[htp!]
\begin{tabular}{ccc}
\includegraphics[width=5.55cm]{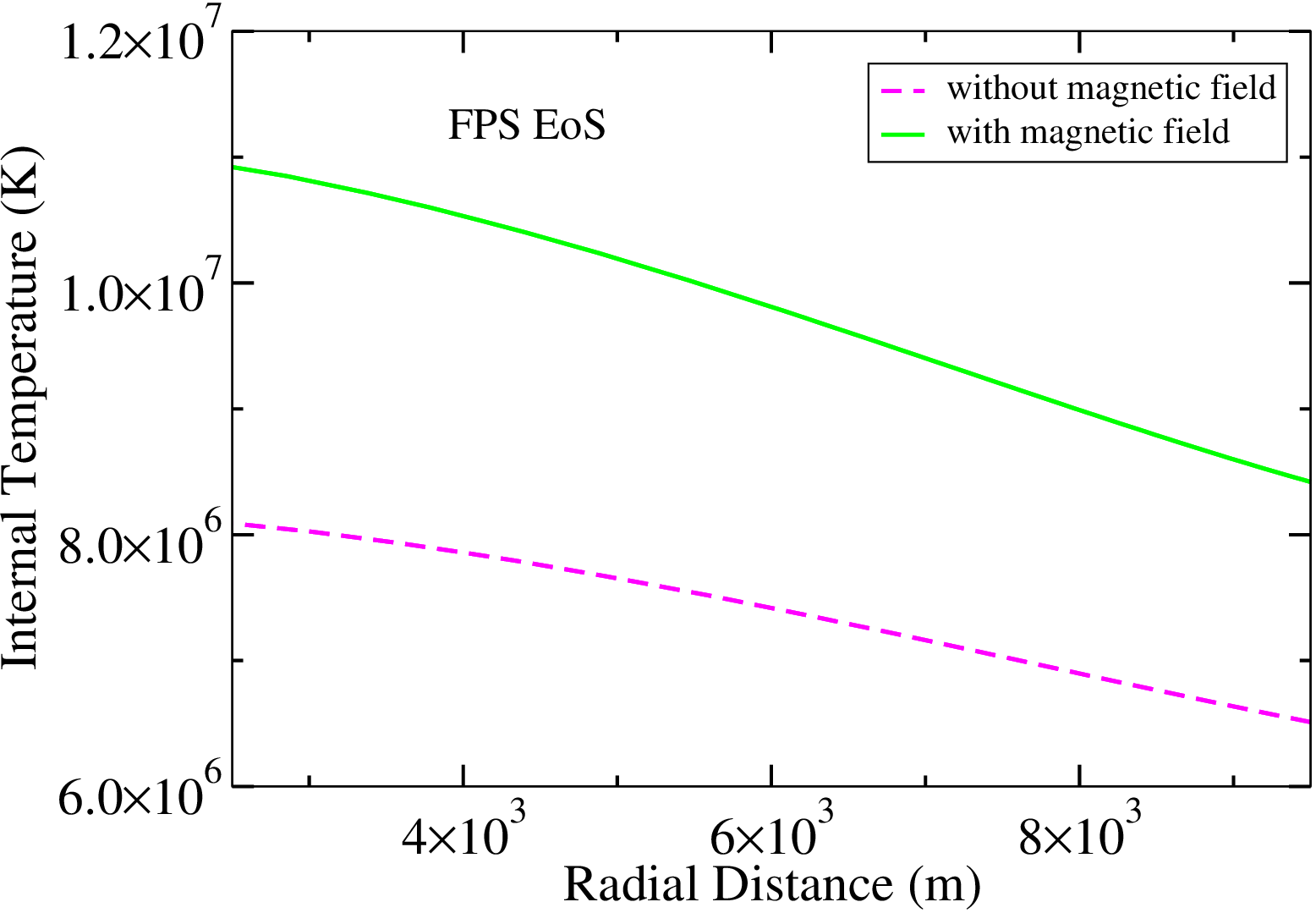}\ & \includegraphics[width=5.55cm]{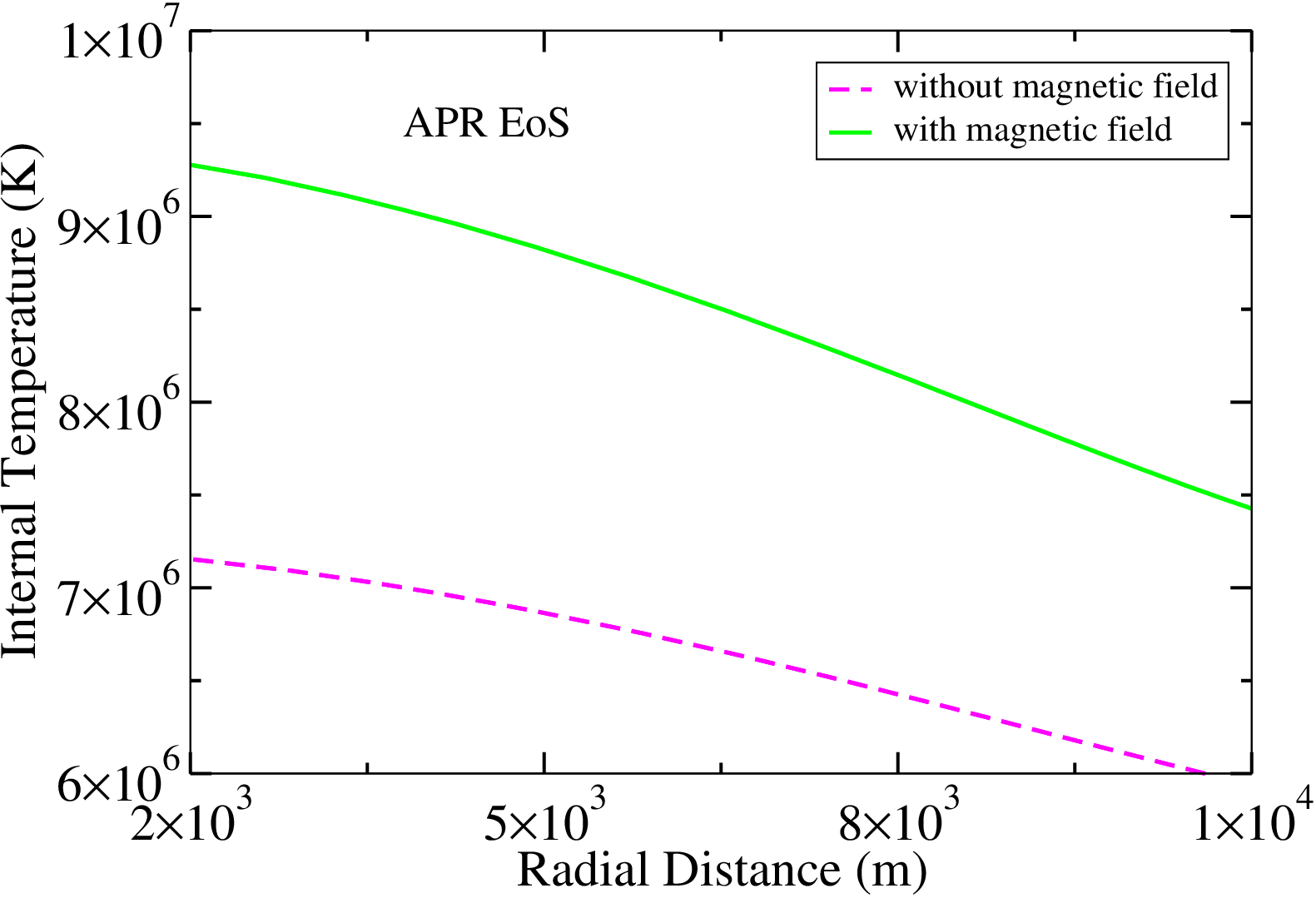}\ & \includegraphics[width=5.55cm]{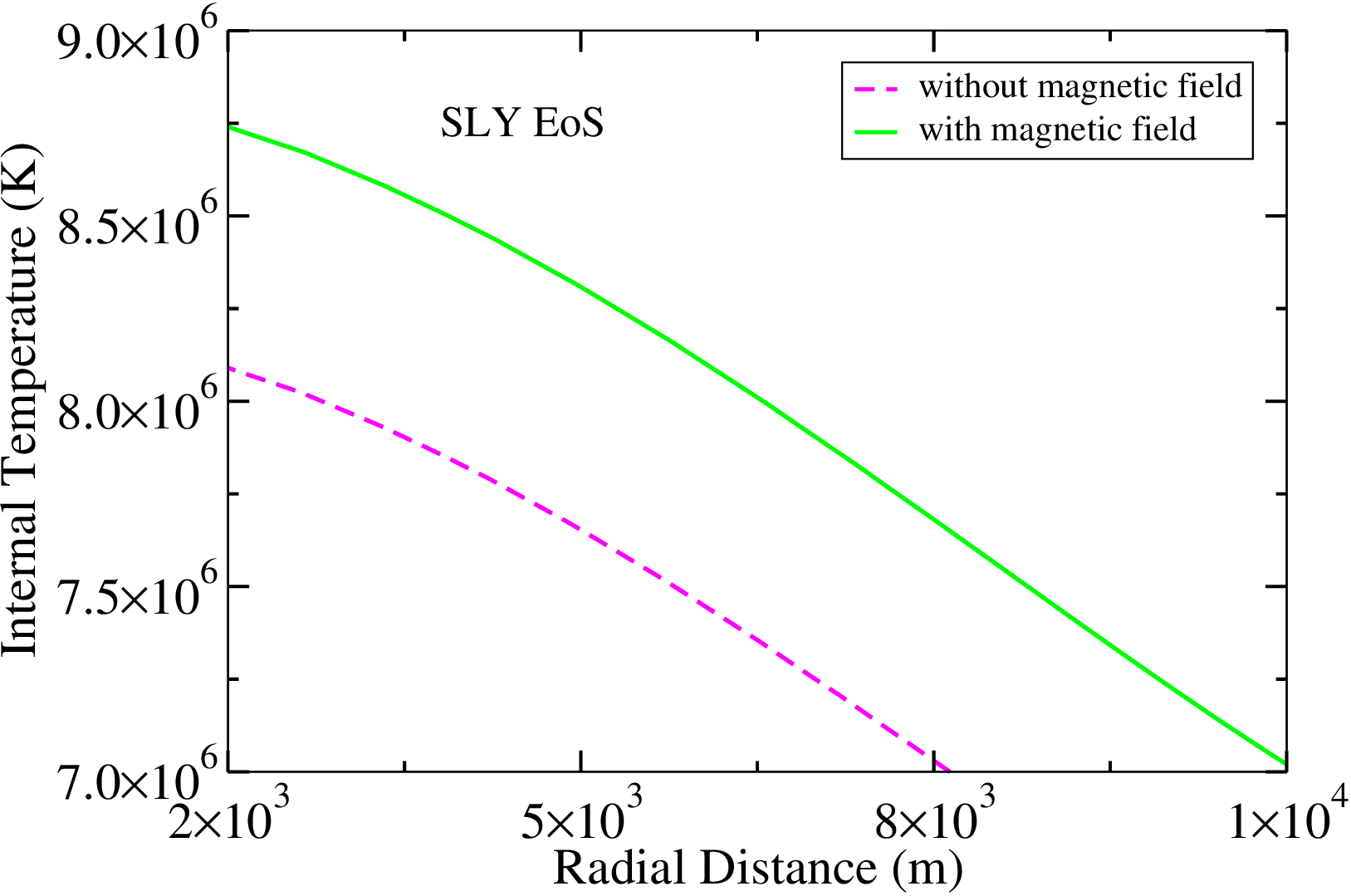} 
\end{tabular}
\caption{The internal temperature distribution versus radial distance (R) for without and with magnetic field at axion mass of 15 meV.} 
\label{fig:nstr2} 
\end{figure*}
%%%%%%%%%%%%%%%%%%%%%%%%%%%%%%%%%%%%%%%
Figure (\ref{fig:nstr2}) presents a distribution of the temperature of interiors versus radial distance at a fixed value of axion mass of 15 meV. The curve for a magnetic field dominates over the corresponding case of without magnetic field at all values of distances.
%%%%%%%%%%%%%%%%%%%%%%%%%%%%%%%%%%%%%%%%%
\begin{figure*}[htp!]
\begin{tabular}{ccc}
\includegraphics[width=5.55cm]{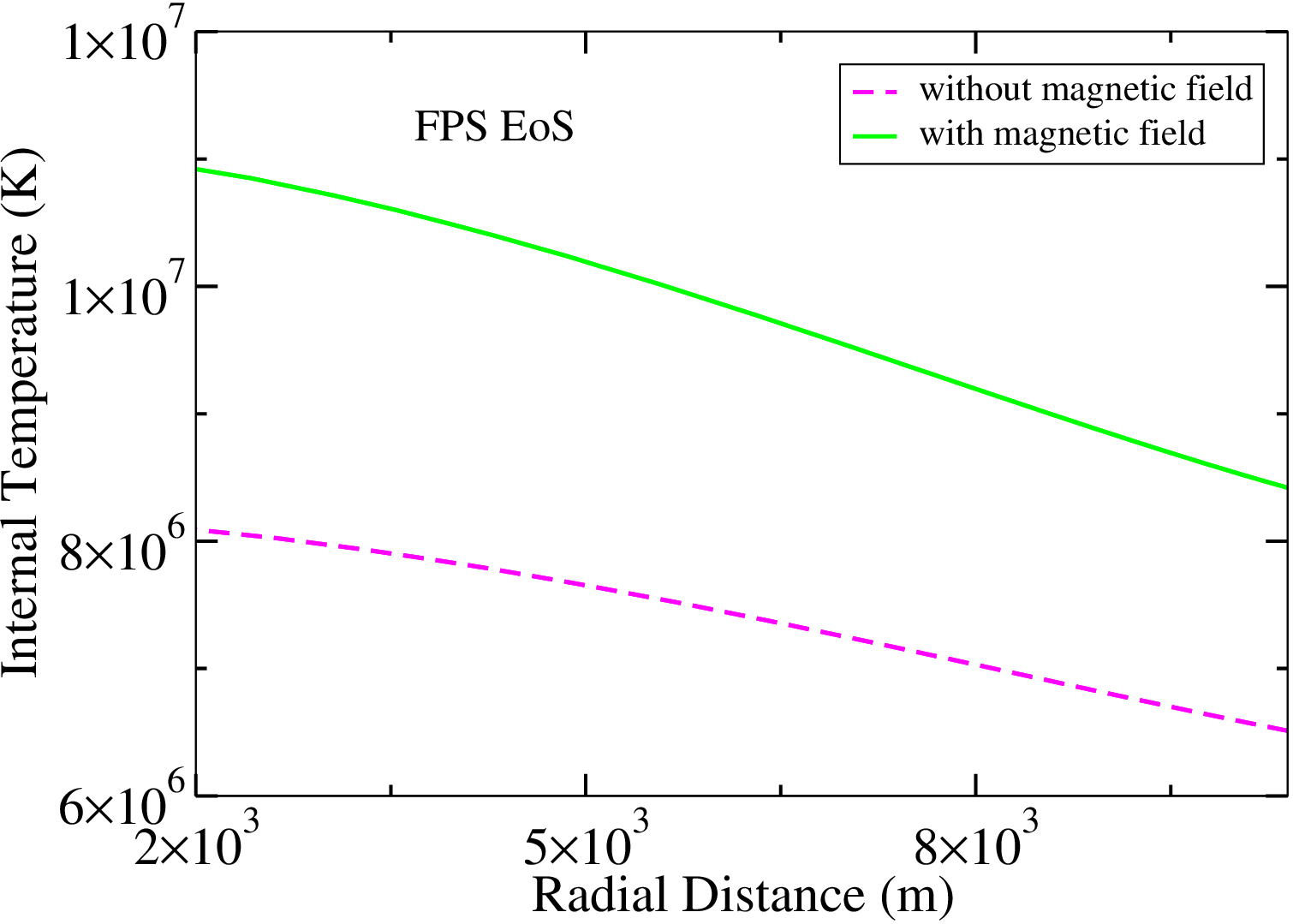}\ & \includegraphics[width=5.55cm]{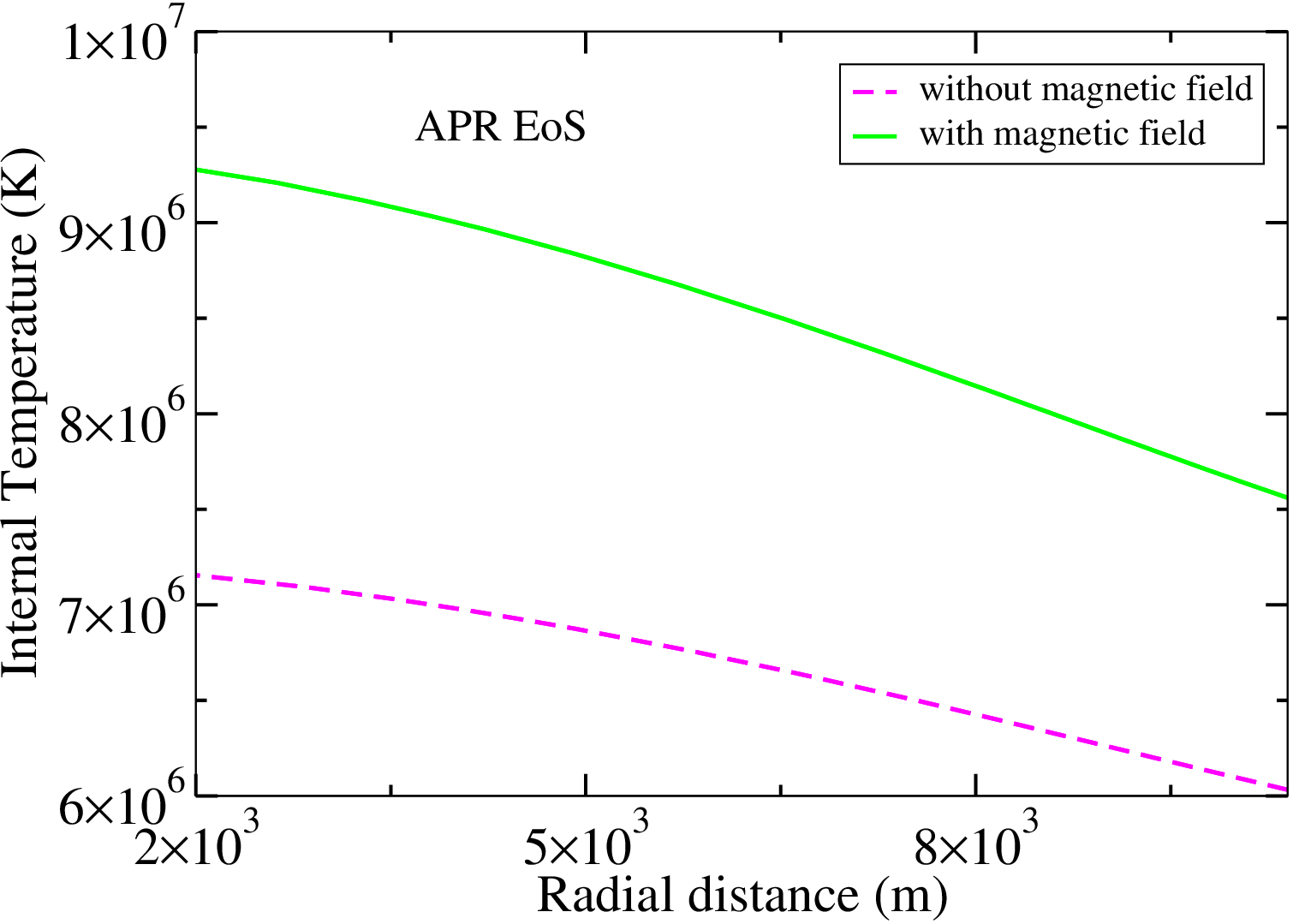}\ & \includegraphics[width=5.55cm]{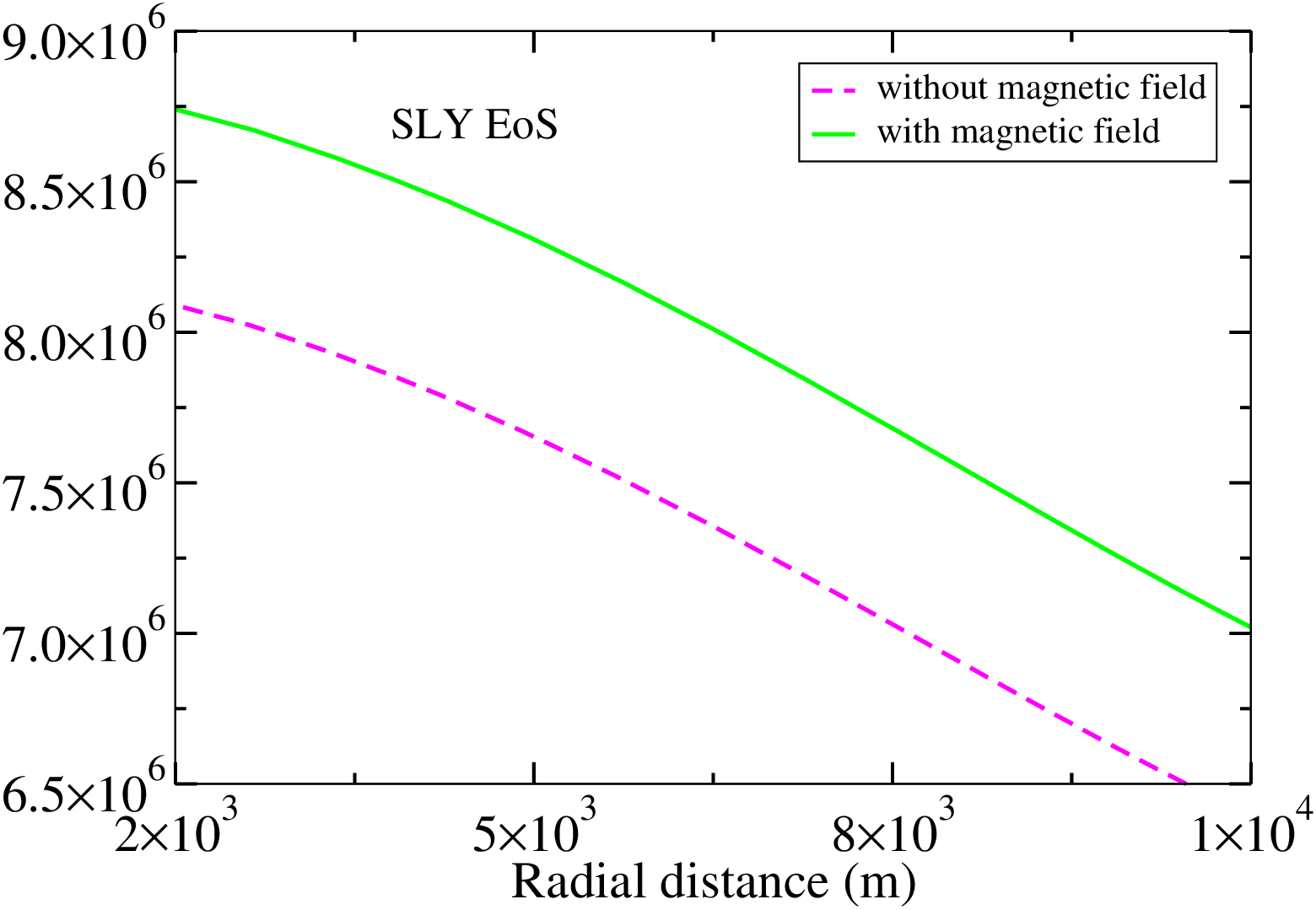} 
\end{tabular}
The internal temperature distribution versus radial distance (R) for without and with magnetic field at axion mass of 20 meV. 
\label{fig:nstr3} 
\end{figure*}
%%%%%%%%%%%%%%%%%%%%%%%%%%%%%%%%%%%%%%%%
Figure (\ref{fig:nstr3}) presents a variation of the internal temperature as a function of radial distance from the centre of NS at an axion mass of 20 meV. The curve of the internal temperature in the presence of a magnetic field remains higher in the case of a magnetic field at all distances. A significant difference can be seen at the early distances up to $3$ km. As we move from the center towards the surface of the NS, the percentage departures for the case with and without a magnetic field decreases, and the separation becomes negligible beyond the distance of $10$ km.

%%%%%%%%%%%%%%%%%%%%%%%%%%%%%%%%%%%%%%%%
 
%%%%%%%%%%%%%%%%%%%%%%%%%%%%%%% luminosity %%%%%%%%%%%%%%%%%%%%%%%%%%%%%%%
%%%%%%%%%%%%%%%%%%%%%%%%%%%%%%%%%%%%%%%%%%%%%%%%%%%%%%%%%%%%%%%%%%%%%%%%%%%%%%%%%%%%%%%%%%%%%%%%%%
\begin{figure*}[htp!]
\begin{tabular}{ccc}
\includegraphics[width=5.55cm]{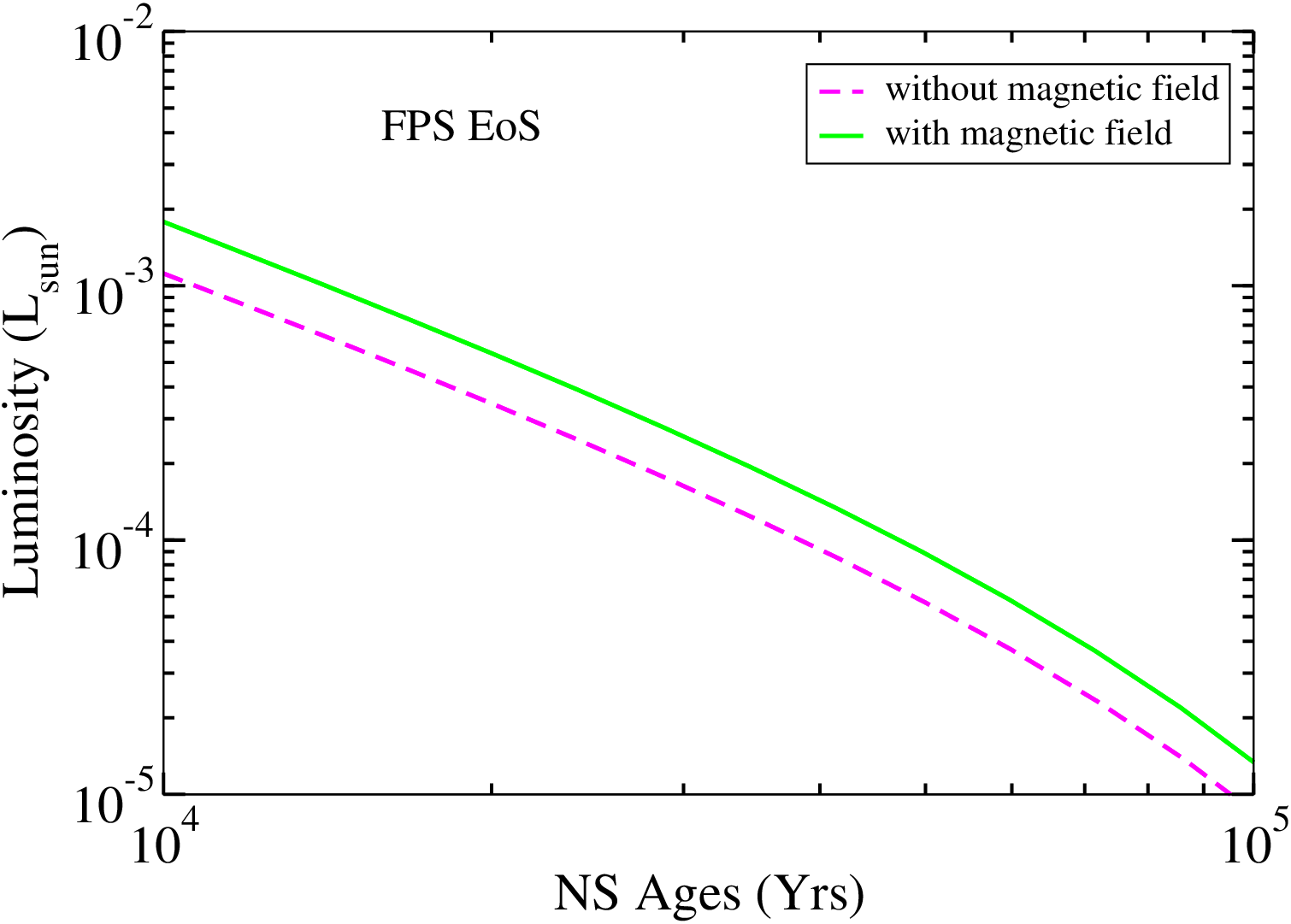}\ & \includegraphics[width=5.55cm]{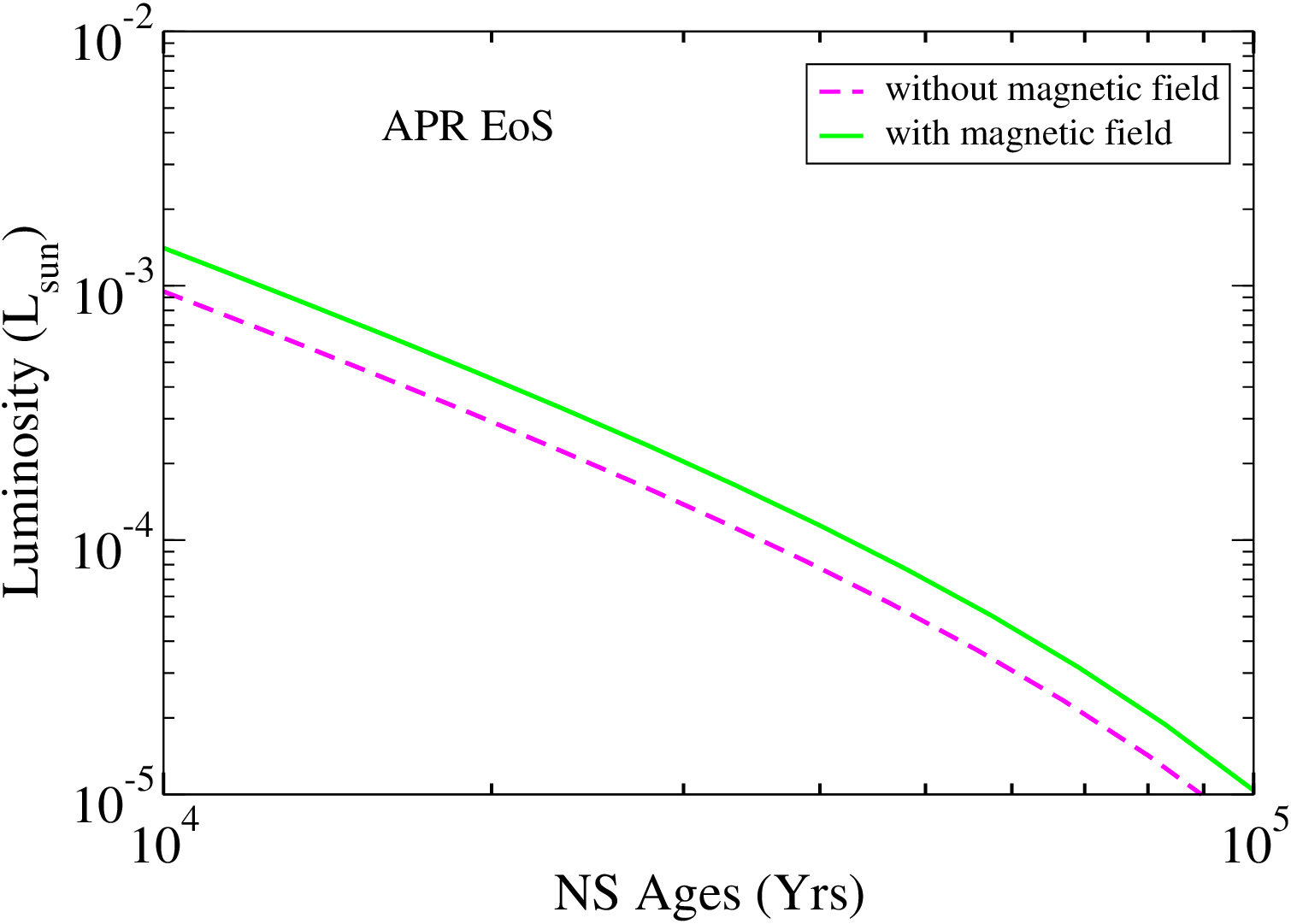}\ & \includegraphics[width=5.55cm]{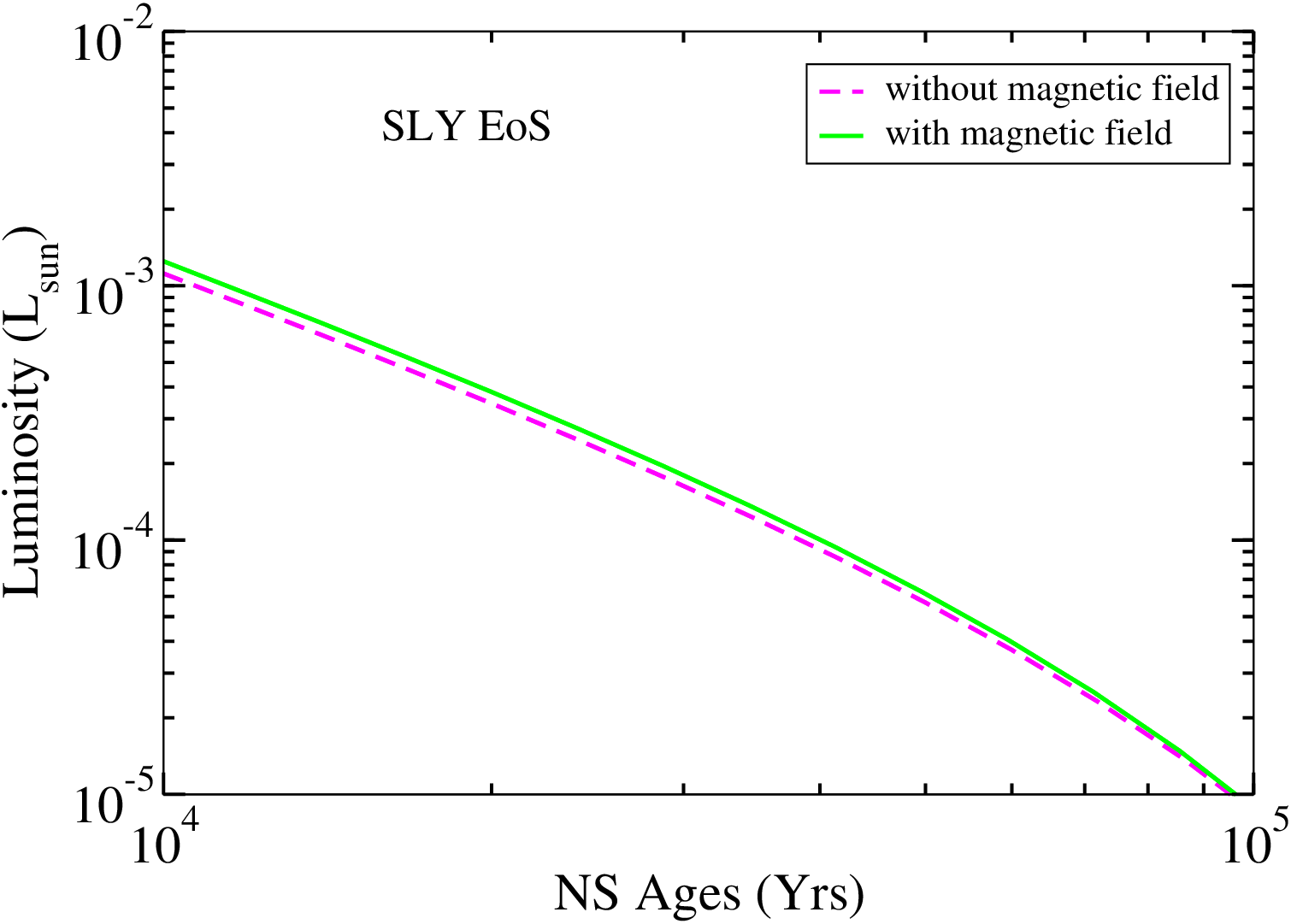} 
\end{tabular}
\caption{The distribution of luminosity for different characteristic ages NS for without and with magnetic field at axion mass of 10 meV.}
\label{fig:nstr4} 
\end{figure*}
%%%%%%%%%%%%%%%%%%%%%%%%%%%%%%%%%%%%%%%
Figure (\ref{fig:nstr4}) depicts the luminosity of axions at an axion mass of 10 meV, comparing the cases of without and with a magnetic field. A constant separation is visible between the ranges of characteristic age for $10^4$ to $10^5$ yrs. The luminosity of axions in the presence of a magnetic field always remains higher than without a magnetic field at all values of characteristic ages. 
%%%%%%%%%%%%%%%%%%%%%%%%%%%%%%%%%%%%%%
\begin{figure*}[htp!]
\begin{tabular}{ccc}
\includegraphics[width=5.55cm]{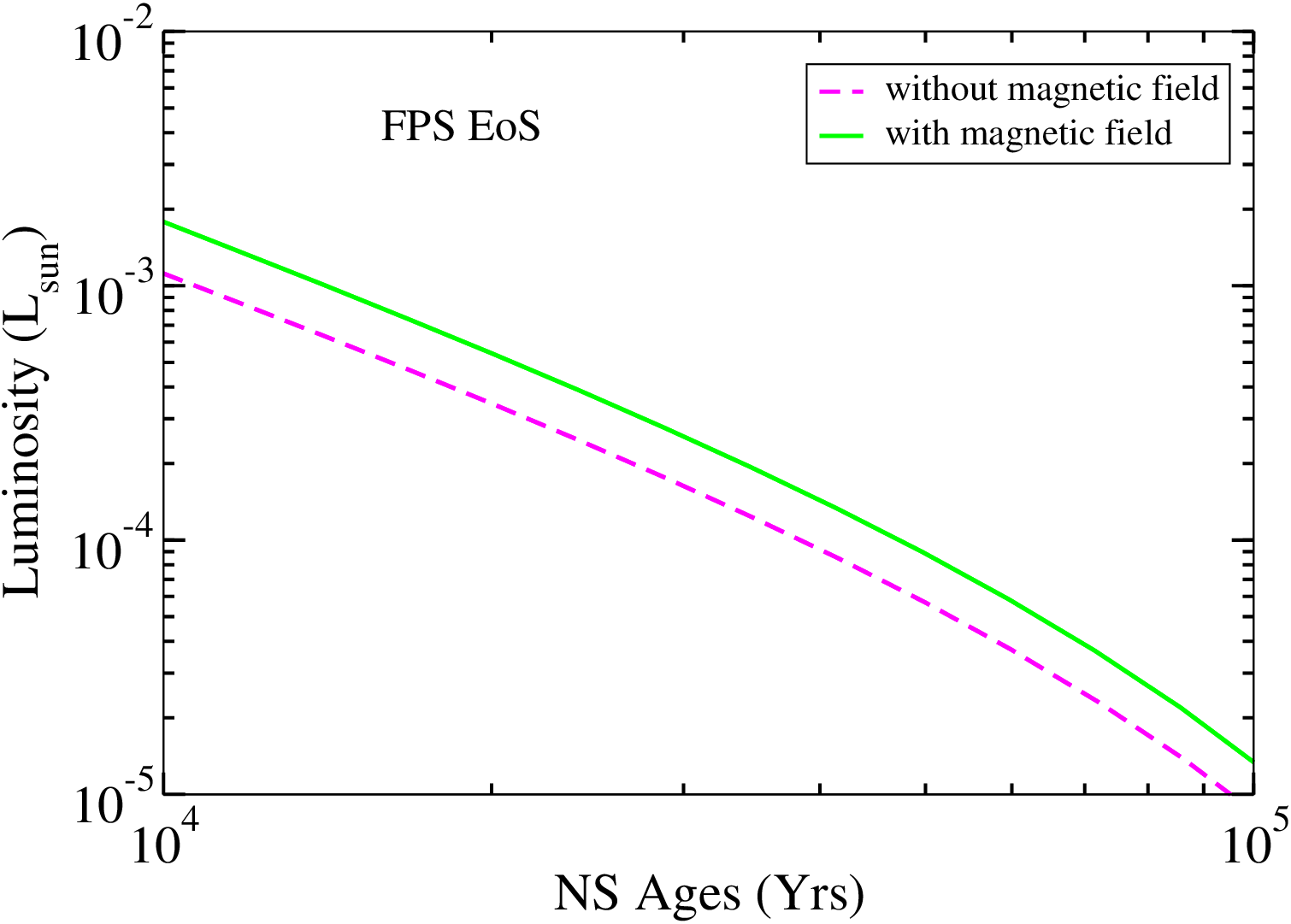}\ & \includegraphics[width=5.55cm]{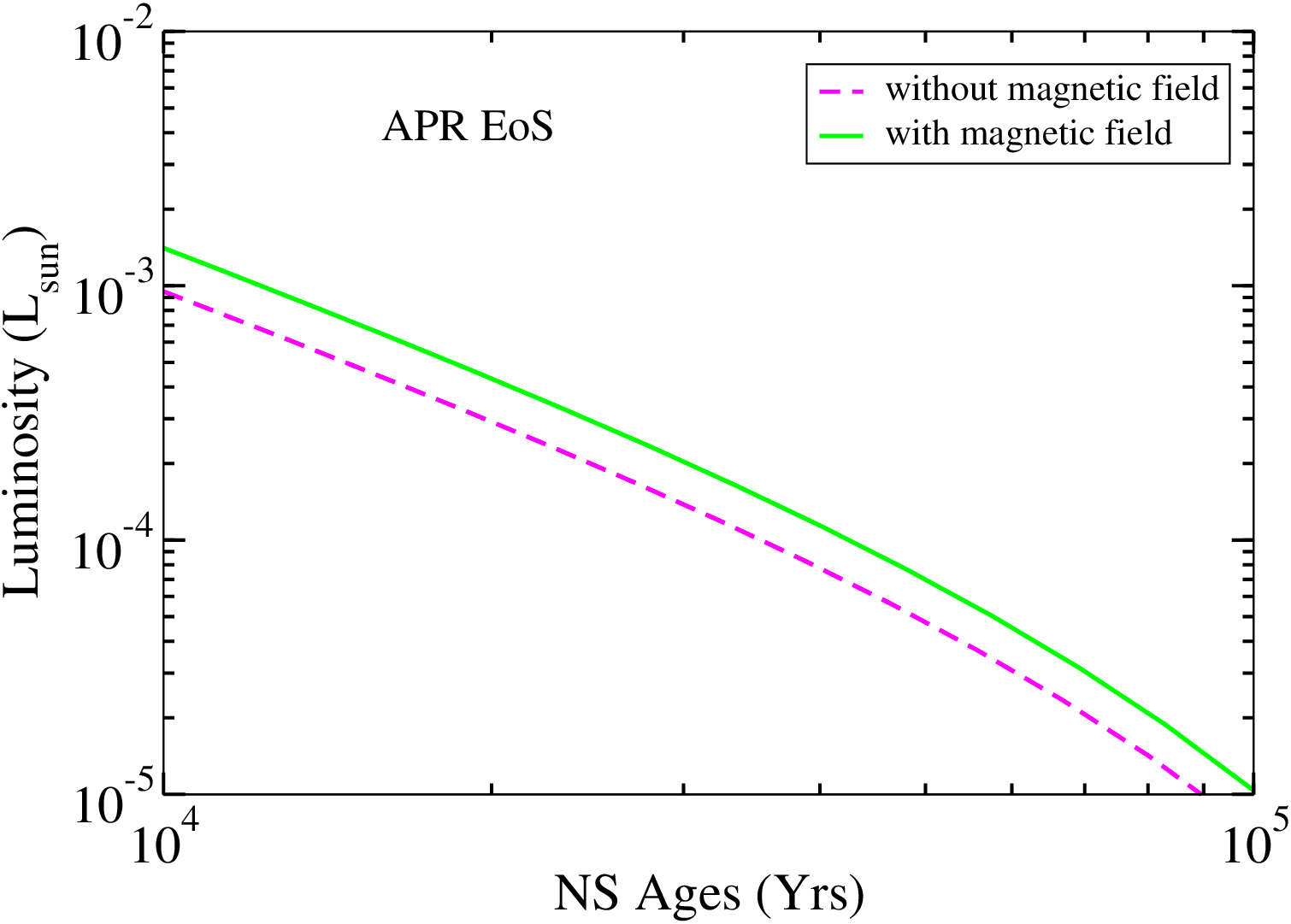}\ & \includegraphics[width=5.55cm]{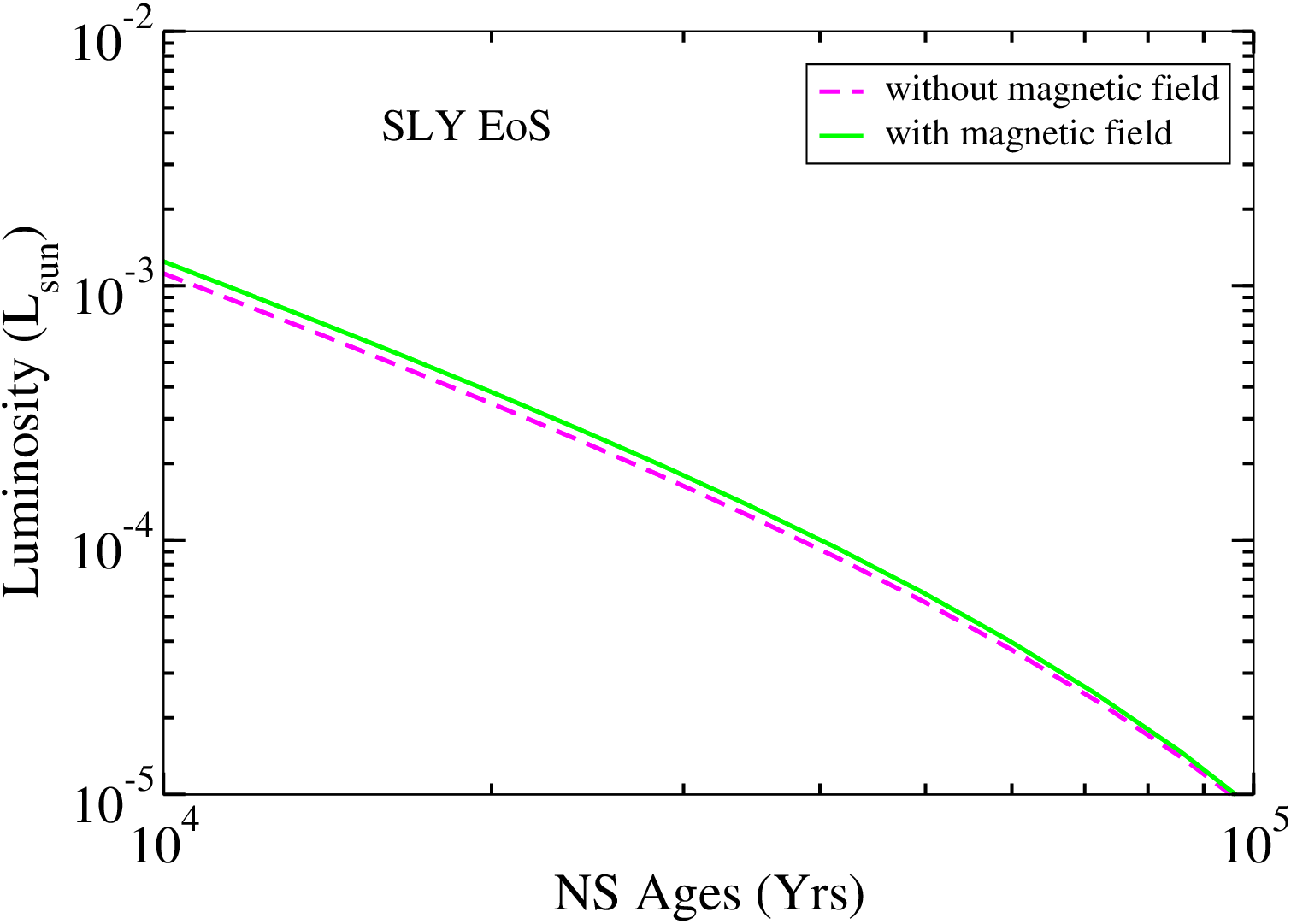} 
\end{tabular}
\caption{The distribution of luminosity for different characteristic ages NS for without and with magnetic field at axion mass of 15 meV.} 
\label{fig:nstr5} 
\end{figure*}
%%%%%%%%%%%%%%%%%%%%%%%%%%%%%%%%%%%%%%%
Figure (\ref{fig:nstr5}) depicts the luminosity of axions at an axion mass of 15 meV, comparing the cases without and with a magnetic field. A smaller separation of axion luminosity is observed for the SLY EoS as compared to the APR and FPS EoS. The curve in the presence of a magnetic field dominates over all possible characteristic ages of NS.
%%%%%%%%%%%%%%%%%%%%%%%%%%%%%%%%%%%%%%%
\begin{figure*}[htp!]
\begin{tabular}{ccc}
\includegraphics[width=5.55cm]{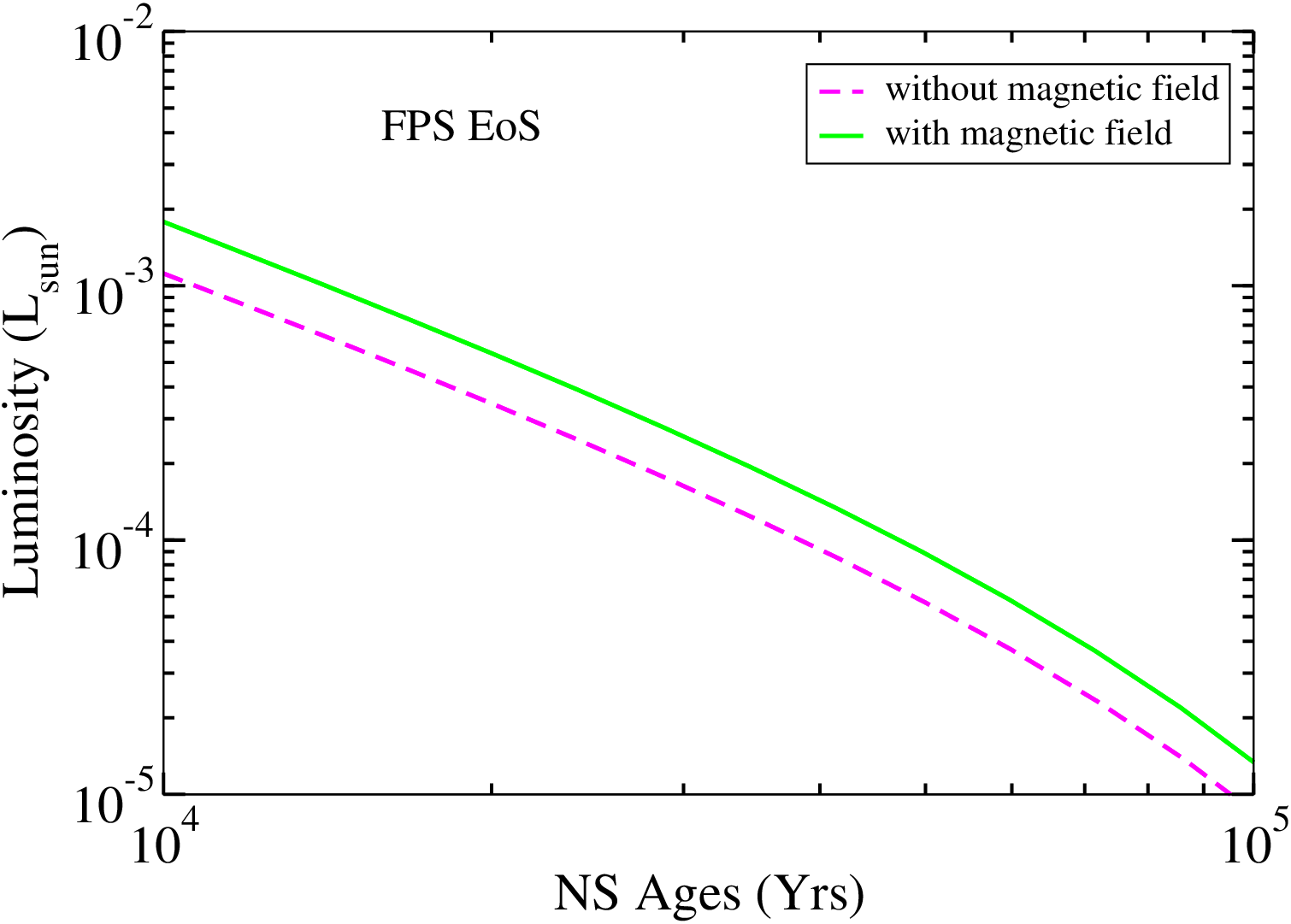}\ & \includegraphics[width=5.55cm]{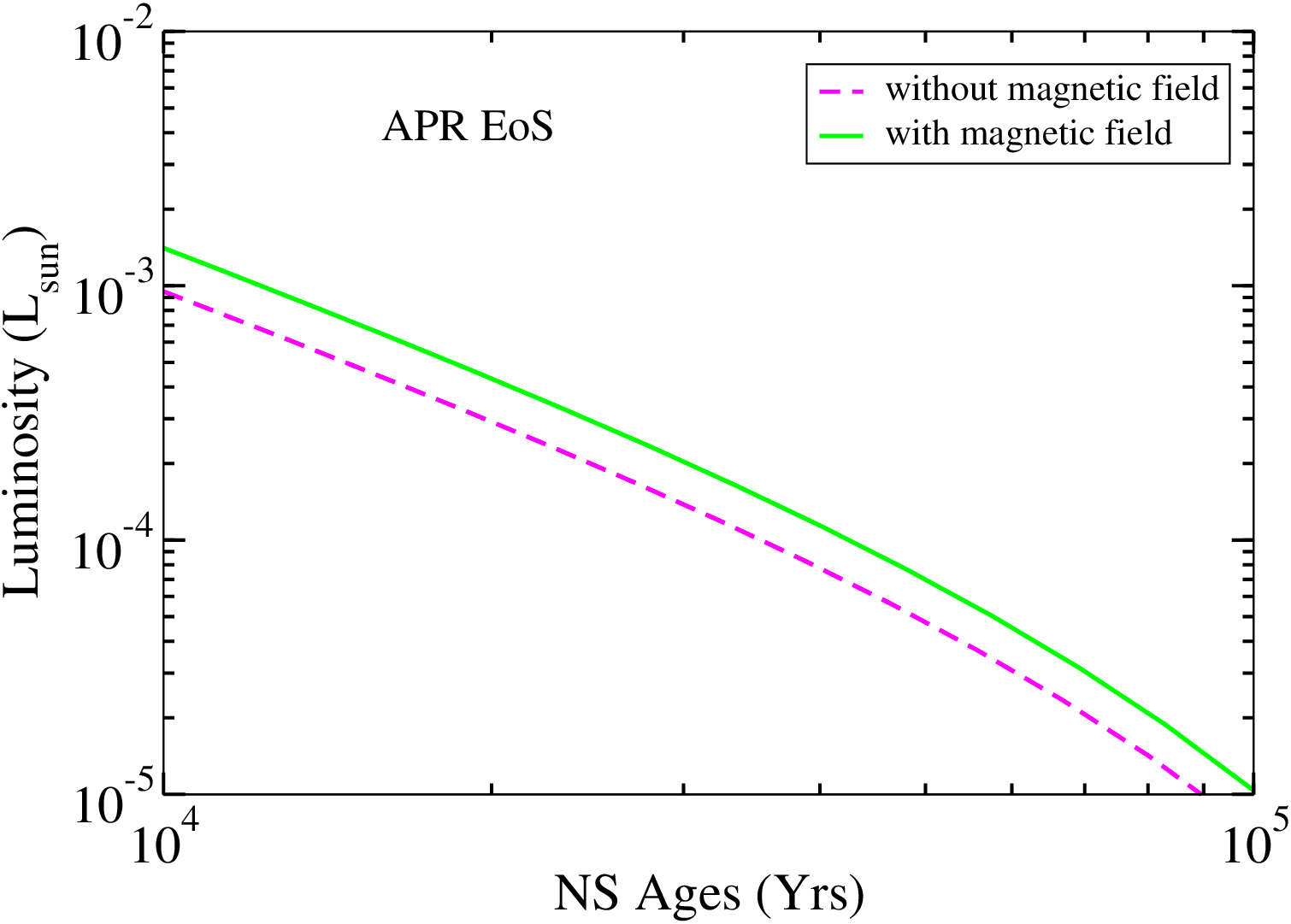}\ & \includegraphics[width=5.55cm]{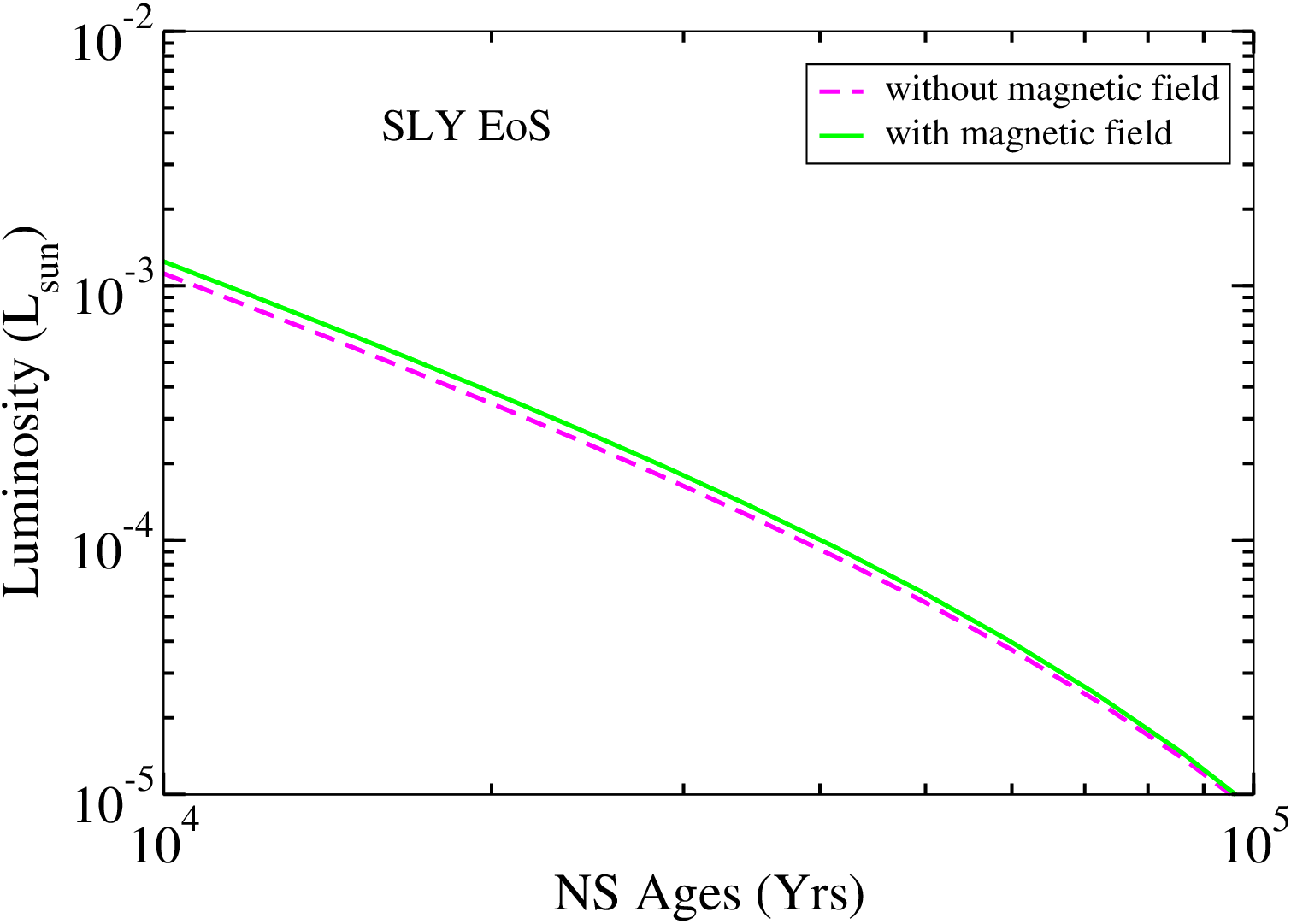} 
\end{tabular}
The distribution of luminosity for different characteristic ages NS for without and with magnetic field at axion mass of 20 meV.
\label{fig:nstr6} 
\end{figure*}
%%%%%%%%%%%%%%%%%%%%%%%%%%%%%%%%%%%%%%%%%
Figure (\ref{fig:nstr6}) shows the variation of the luminosity of axions as a function of the characteristic age for the APR, FPS, and SLY EoS at an axion mass of 20 meV. With increasing ages, the luminosity of axions in the presence of a magnetic field remains higher than in the absence of a magnetic field. Also, the separation between the luminosity of axions (with/without magnetic fields) is minimum for the SLY EoS as compared to the APR and FPS EoS. The effect of the strong fields is more pronounced at greater values of the characteristic ages of NSs.
%%%%%%%%%%%%%%%%%%%%%%%%%%%%%%%%%%%%% energy spectrum %%%%%%%%%%%%%%%%%%%%%%%%
\begin{figure*}[htp!]
\begin{tabular}{ccc}
\includegraphics[width=5.55cm]{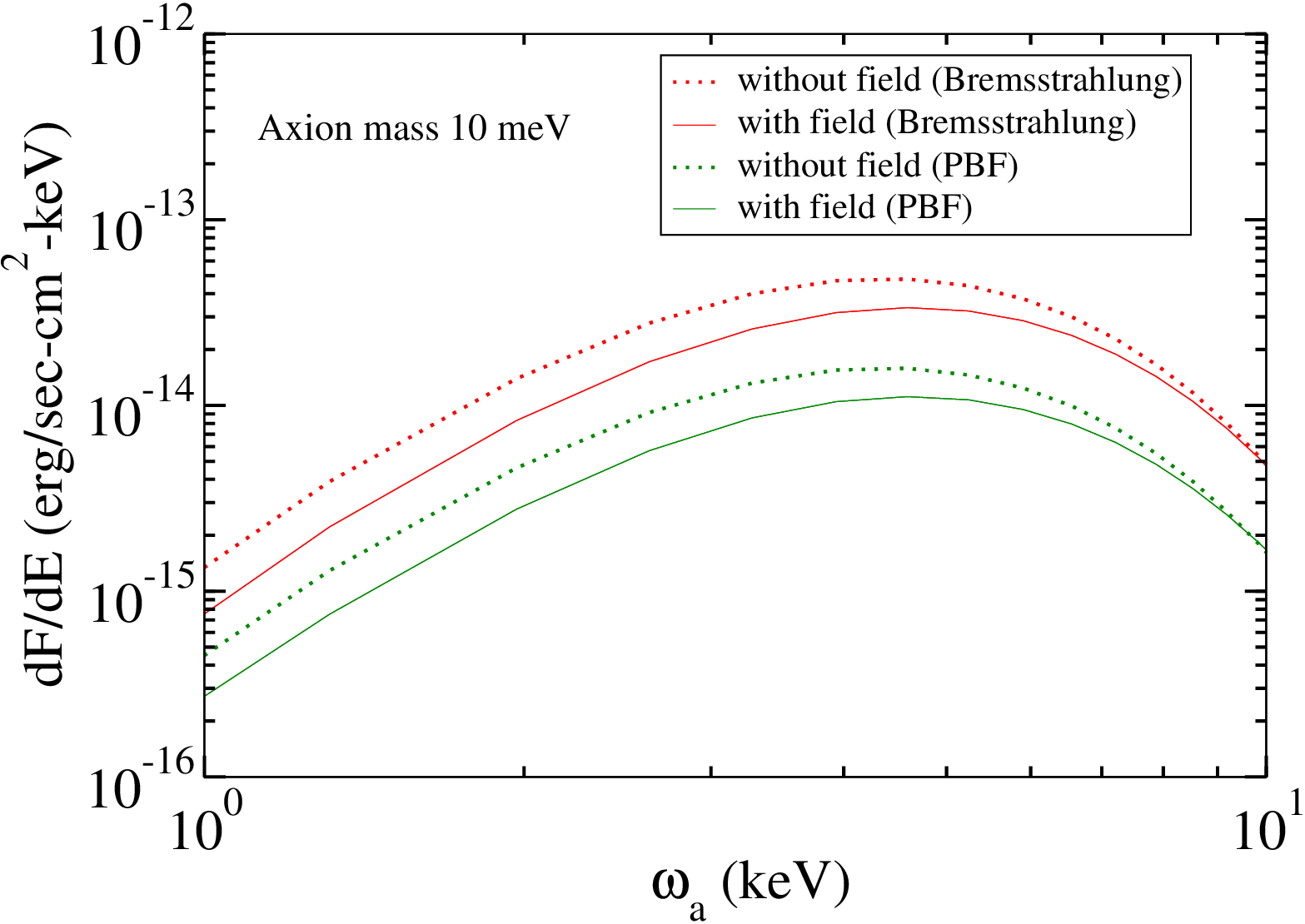}\ & \includegraphics[width=5.55cm]{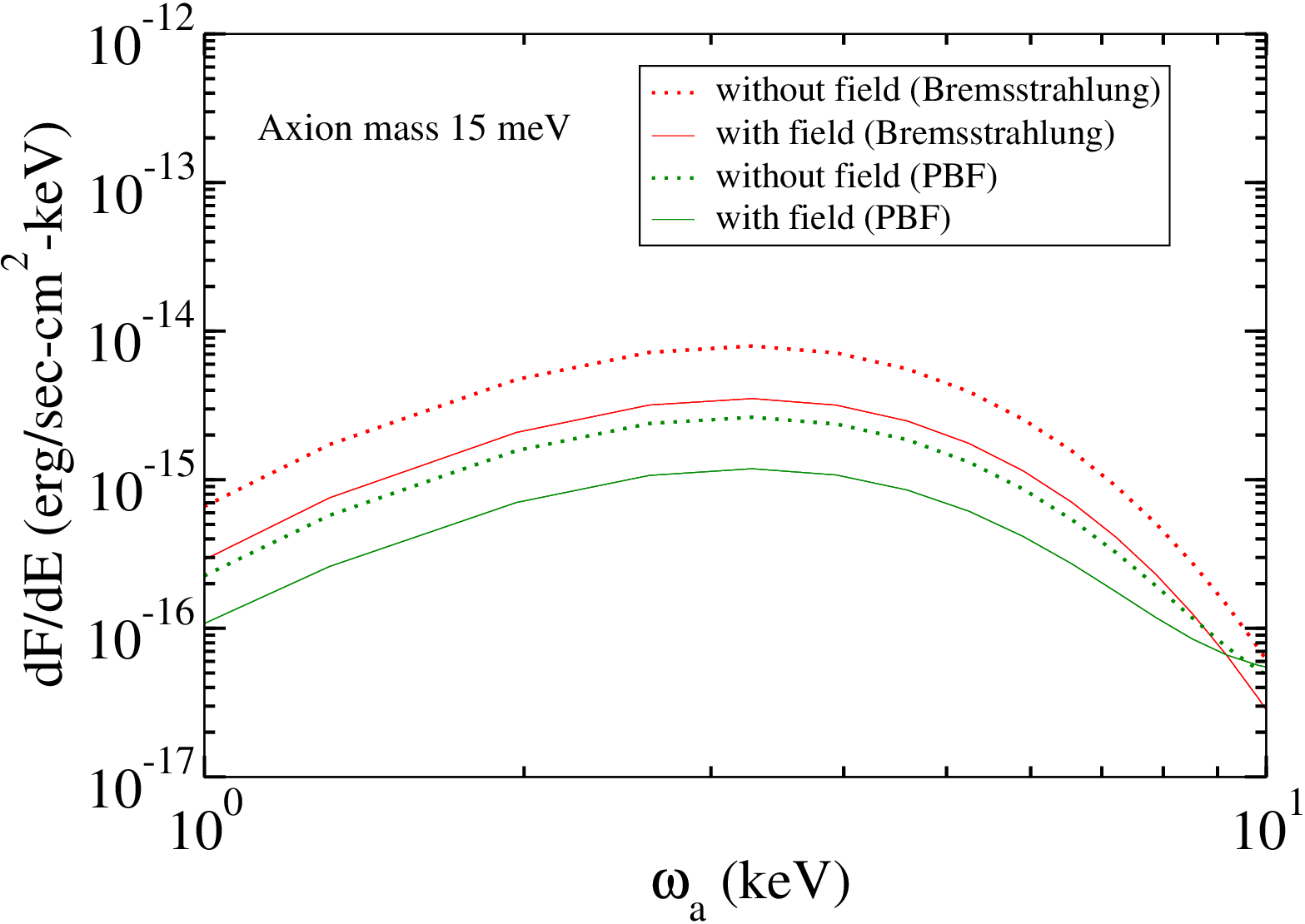}\ & \includegraphics[width=5.55cm]{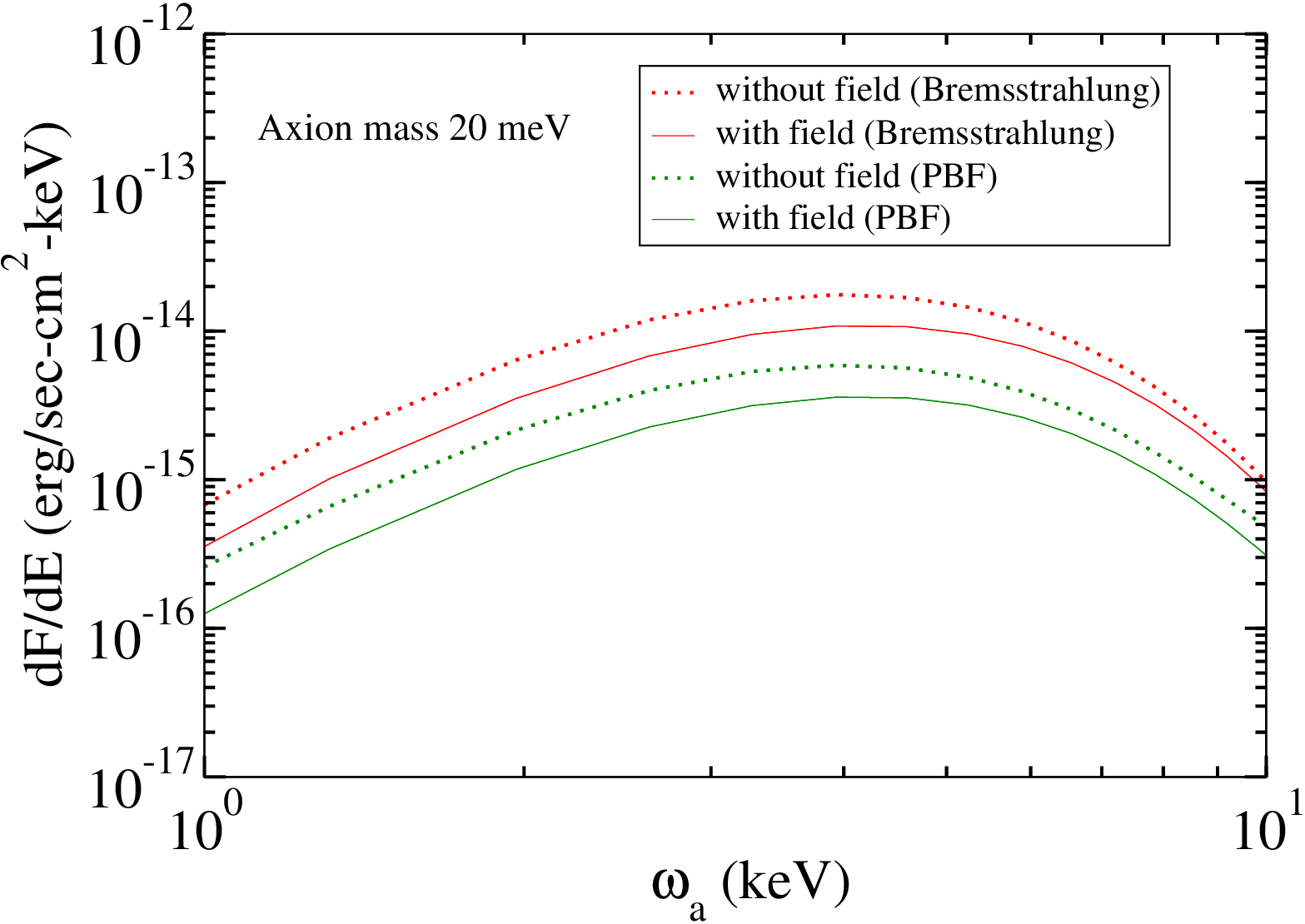} 
\end{tabular}
\caption{The energy spectrum for different masses of axions versus axion energies comparing with and without magnetic field of FPS EoS.} 
\label{fig:nstr7} 
\end{figure*}
Figure (\ref{fig:nstr7}) presents a variation of the energy spectrum of axions for several ranges of axion energies (1-10 keV) at axion masses of 10 meV, 15 meV, and 20 meV for PSR J1357-6429 NS. In the Bremsstrahlung process only, the largest difference between the energy spectra of axions in the case with and without a magnetic field is seen at 15 meV, and eventually the separation decreases with increase in axion energy. The separation becomes negligible for 10 meV and 20 meV at larger values of axion energies. Finally, both curves converges beyond 10 keV axion energies. For the PBF process, a noteworthy separation is observed for an axion mass of 15 meV as compared to 10 meV and 20 meV, with and without a magnetic field. For lower values of the axion energies, the curve in the absence of magnetic fields is more dominant over the magnetic field. A constant departure can be seen up to 7 keV of axion energy; after, it decreases beyond 10 keV. Specifically, for the lower ranges of the axion energies, the net energy spectrum from the Bremsstrahlung process dominates over the PBF process at all axion masses (10 meV, 15 meV, and 20 meV) occurring within the core of NS. 
%%%%%%%%%%%%%%%%%%%%%%%%%%%%%%%%%%%%%%%photon flux%%%%%%%%%%%%%%%%%%%%%%%%%%%%%%%%%%%%%
\begin{figure*}[htp!]
\begin{tabular}{ccc}
\includegraphics[width=5.55cm]{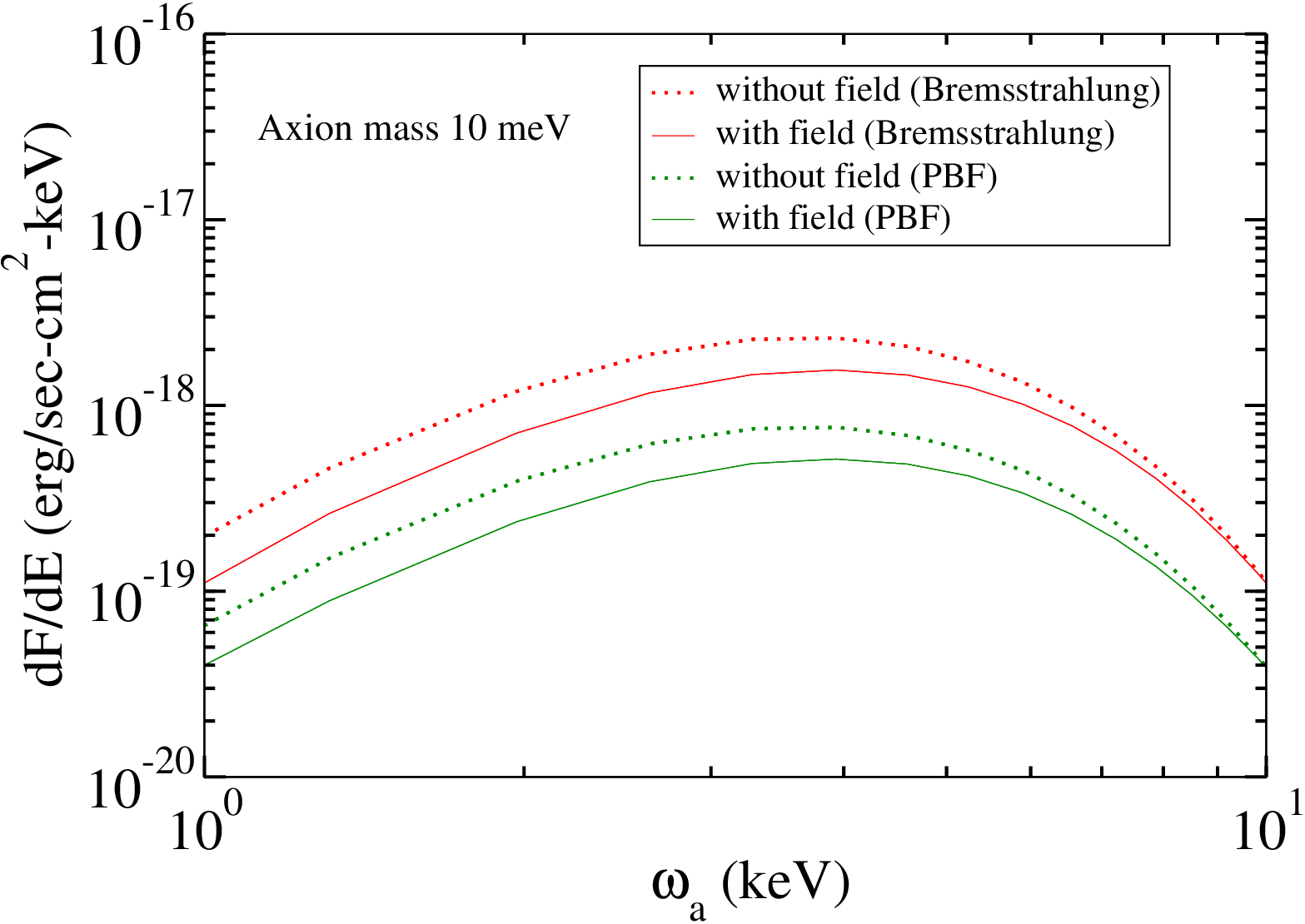}\ & \includegraphics[width=5.55cm]{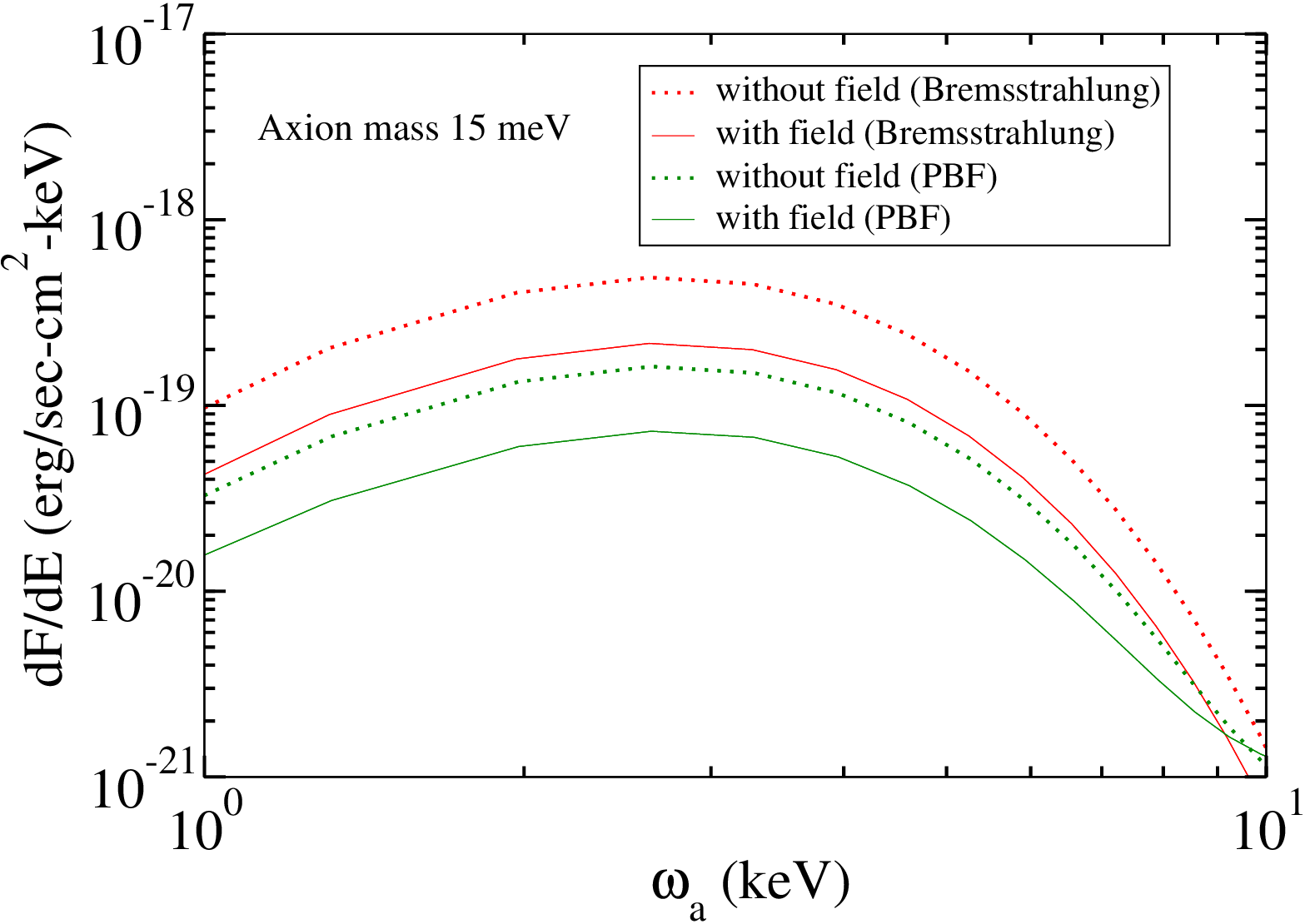}\ & \includegraphics[width=5.55cm]{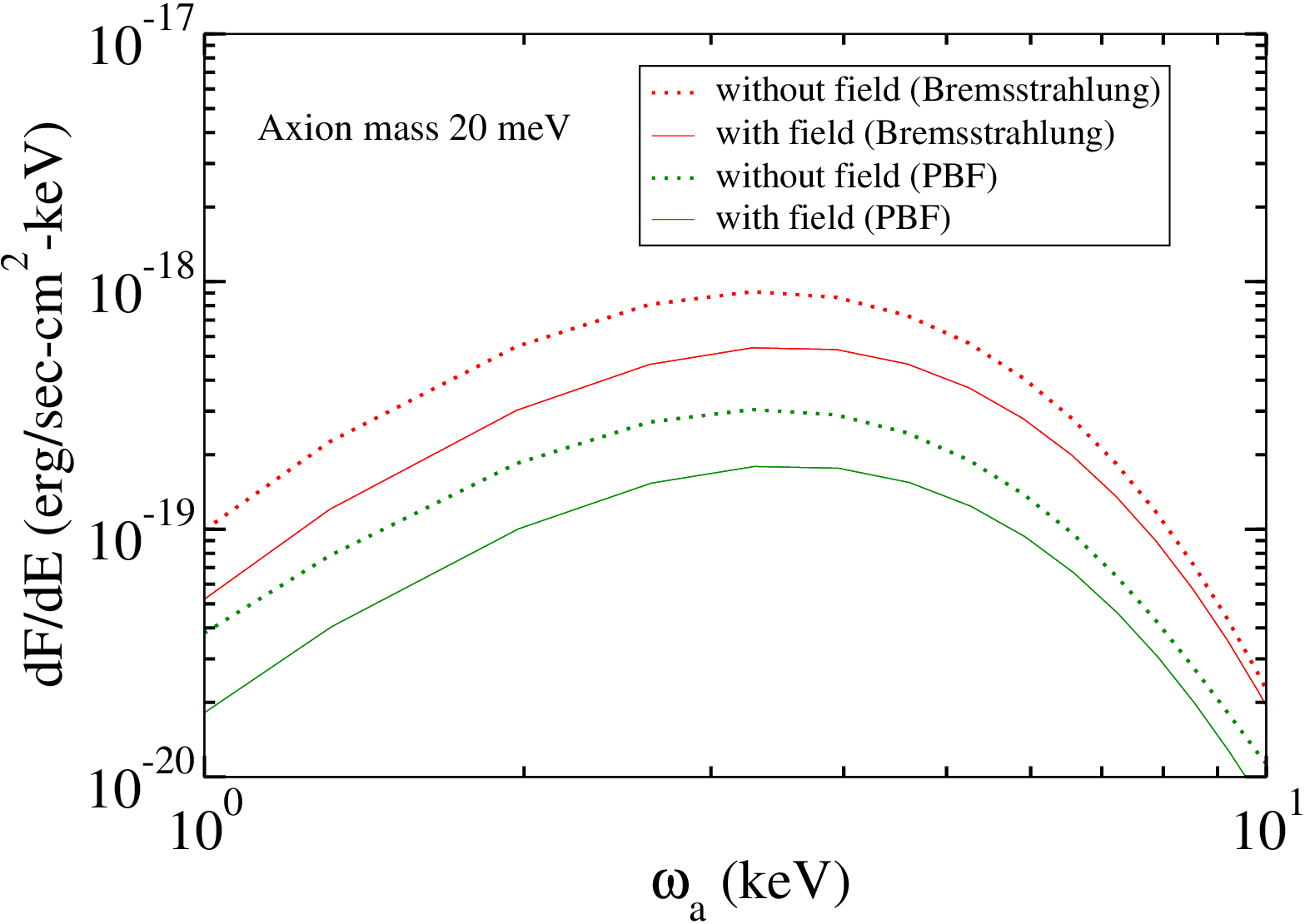} 
\end{tabular}
\caption{The axion-converted-to-photon flux for different masses of axions versus axion energies comparing with and without magnetic field of FPS EoS.} 
\label{fig:nstr8} 
\end{figure*}
Figure (\ref{fig:nstr8}) presents the distribution of axion-converted-to-photon flux as a function of axion energies (1 keV - 10 keV) within the axion masses 10 meV, 15 meV, and 20 meV. We present a comparative study for the PBF and the Bremsstrahlung process only in the core of PSR J1357-6429 NS. For the Bremsstrahlung process, a smaller separation of photon flux in the cases of presence and absence of a magnetic field is observed at an axion mass of 10 meV than at 15 meV and 20 meV. The curve for without a magnetic field dominates over the case of a magnetic field at all values of the axion energies. For a no-magnetic field, the decrease in photon flux is at a faster rate for 10 meV and 20 meV than for 15 meV. For the PBF process, a large departure is visible in the photon flux at an axion mass of 15 meV and 20 meV within the axion energies ranging from 1 keV to 4 keV, in comparison to 10 meV mass. A sudden overlap in the distribution of photon flux is observed at 15 meV at a 10 keV axion energy. Beyond this, the separation decreases and both the curves overlaps for axion energies greater than 10 keV. The net flux of axion-to-photon conversion originating from the Bremsstrahlung process dominates over the PBF process for lesser axion energy for $\sim$ 8 keV.
\subsection{Observational analysis and Detectability}
The predicted photon emission from magnetised NS may provide an interesting observational signature related to the photon signals. The detectability of such signals depends on several variables, such as the luminosities of photons, the distribution of the spectrum, radial distances, interstellar absorption, and the sensitivity of the instrument. Other requirement is detailed comparison of the calculated photon flux and its distribution within the sensitivity of the instrument. Such instrument-specific investigation is beyond the current study and might be useful in enhancing the understanding of theoretical implications with the observational predictions.
\section{Conclusion and Future Outlook} 
\label{conc}
The emission properties of the magnetised NSs are related to certain factors, including gravitational effects, radiation from distant stellar objects, strong magnetic fields, and several observations from astrophysical objects. In the present manuscript, we calculate the influence of strong magnetic fields on the different axion masses of 10 meV, 15 meV, and 20 meV by employing APR, FPS, and SLY EoS. Our prime focus is on PSR J1357-6429 NS, a non M7 star and study its properties. This is in continuation of our several prior work~\cite{yadav2022emission,yadav2024thermal,yadav2025conversion,yadav2025magnetic,yadav2024x}. The difference relies on the detailed influence of axion masses and EoS on several aspects of cooling and detailed mechanism for the energy spectrum and photon flux of the star. We emphasize on axions production mechanism only in the core of NS. The cooling rate and luminosity are quite sensitive to axion masses of 10 meV, 15 meV, and 20 meV, with effect of strong fields more pronounced at 10 meV and 15 meV. The results reveal that axion mass is sensitive to the employed EoS and the specific NS. A minimal increase of the luminosity of axions is observed for the SLY EoS in comparison to the APR and FPS EoS in the presence of a magnetic field. The influence of strong fields on the energy spectrum of axions and axion-converted-to-photon flux is less for all axion masses, 10 meV, 15 meV, and 20 meV. We find that Bremsstrahlung process dominates over PBF process at lower axion energies. It is also dependent on the EoS and consistent with all previously observed results. We conclude the necessity of Magnificent M7 stars provides an excellent environment for better analysis of properties of NS. These findings would develop a strong foundation for the analysis of the astrophysical environment of NS.

\section*{Acknowledgements}

We thank D.Page and M.Buschmann for answering questions regarding the NSCool code. The authors acknowledges the Oriental Institute of Science and Technology, Bhopal, Madhya Pradesh for research facilities and financial assistance. 

\section*{Data Availability Statement}

No Data associated in the manuscript.

 \vskip 1cm
\bibliography{MS_CRSV2}  
%\vskip 5cm
\newpage
\end{document}